\documentclass[12pt]{article}
\usepackage{amsfonts}
\usepackage{amssymb}
\usepackage{amsbsy}
\usepackage{graphics}

\input epsf.sty

\newlength{\TZ}
\newcommand{\BEQ}{\begin{equation}}     % Gleichungen Anfang ..
\newcommand{\BEA}{\begin{eqnarray}}
\newcommand{\EEQ}{\end{equation}}       % .. und Ende
\newcommand{\EEA}{\end{eqnarray}}
\newcommand{\eps}{\varepsilon}          % epsilon
\newcommand{\vph}{\varphi}              % rundes phi
\newcommand{\D}{{\rm d}}                % gerades d fuer Ableitungen
\newcommand{\II}{{\rm i}}               % gerades i fuer komplexe Einheit
\newcommand{\wit}[1]{\widetilde{#1}}    % weite Schlange
\newcommand{\wht}[1]{\widehat{#1}}      % weiter Hut
\renewcommand{\vec}[1]{\boldsymbol{#1}} % Vektoren fettgedruckt

\newcommand{\zeile}[1]{\vskip #1 \baselineskip} % N Zeilen ueberschlagen
\newcommand{\vekz}[2]
     {\mbox{${\begin{array}{c} #1  \\ #2 \end{array}}$}}
\newcommand{\appsection}[2]{\setcounter{equation}{0}\setcounter{subsection}{0}
\section*{Appendix #1. #2}
\renewcommand{\theequation}{#1\arabic{equation}}
              \renewcommand{\thesection}{#1} }
\catcode`\@=11
\def\numberbysection{\@addtoreset{equation}{section}
        \def\theequation{\thesection.\arabic{equation}}}
\numberbysection

\begin{document}

\begin{titlepage}

%{\hfill \tt \today; pr\'eliminaire !}

\vskip 1.5 cm
\begin{center}
{\Large \bf Local scale-invariance and ageing in noisy systems}
\end{center}

\vskip 2.0 cm
\centerline{  {\bf Alan Picone} and {\bf Malte Henkel} }
\vskip 0.5 cm
\centerline {Laboratoire de Physique des 
Mat\'eriaux,\footnote{Laboratoire associ\'e au CNRS UMR 7556} 
Universit\'e Henri Poincar\'e Nancy I,} 
\centerline{ B.P. 239, 
F -- 54506 Vand{\oe}uvre l\`es Nancy Cedex, France}

\begin{abstract}
The influence of the noise on the long-time ageing dynamics of a quenched 
ferromagnetic spin system with a non-conserved order parameter and described 
through a Langevin equation with a thermal noise term and a disordered initial 
state is studied. If the noiseless part of the system is 
Galilei-invariant and scale-invariant with dynamical exponent $z=2$, 
the two-time linear response function is independent of the noise and therefore 
has exactly the form predicted from the local scale-invariance of the noiseless 
part. The two-time correlation function is exactly given in terms of
certain noiseless three- and four-point response functions. An explicit scaling 
form of the two-time autocorrelation function follows. For disordered
initial states, local scale-invariance is sufficient for the equality of
the autocorrelation and autoresponse exponents in phase-ordering kinetics. 
The results for the scaling functions are confirmed through tests in the 
kinetic spherical model, the spin-wave approximation of the XY model, 
the critical voter model and the free random walk. 
\end{abstract}
\zeile{2} \noindent
\end{titlepage}

%%%%%%%%%%%%%%%%%%%%%%%%%%%%%%%%%%%%%%%%%%%%%%%%%%%%%%%%%%%%%%%%%%%%%%%%%%%%%%%%
\section{Introduction}
%%%%%%%%%%%%%%%%%%%%%%%%%%%%%%%%%%%%%%%%%%%%%%%%%%%%%%%%%%%%%%%%%%%%%%%%%%%%%%%%

The study of ageing phenomena as they are known to occur in glassy and
non-glassy systems presents one of the great challenges in current research
into strongly coupled many-body systems far from thermal equilibrium. 
A common example of this kind of system is obtained as follows. Consider a
magnet at a high-temperature initial state before quenching it to a 
final temperature $T$ at or below its critical temperature $T_c>0$. 
Then the temporal evolution of the system with $T$ fixed is studied. 
A key insight has been the observation that many of the apparently erratic
and history-dependent properties of such systems can be organized in terms of
a simple scaling picture \cite{Stru78}. Underlying this phenomenological picture
is the idea that the ageing phenomenon and the related slow evolution of the
macroscopic observables comes from the slow motion of the domain walls which
separate the competing correlated clusters. The domains are of a typical 
time-dependent size with length-scale $L(t)$, 
see \cite{Bray94,Bray00,Bouc00,Godr02,Cugl02,Cris03} 
for reviews. In recent years, much work has been performed on the
ageing phenomena of simple ferromagnetic systems, in the hope that these
systems might offer insight useful for the 
refined study also of ageing glassy materials. It has turned out that ageing
is more fully revealed in two-time observables, such as the two-time
(auto-)correlation function $C(t,s)$ or the two-time linear 
(auto-)response function $R(t,s)$ defined as
\BEQ \label{1:gl:CR}
C(t,s) := \langle \phi(t) \phi(s) \rangle \;\; , \;\; 
R(t,s) := \left.\frac{\delta \langle \phi(t)\rangle}{\delta h(s)}\right|_{h=0}
\EEQ
where $\phi(t)$ denotes the time-dependent order parameter, $h(s)$ is the
time-dependent conjugate magnetic field, $t$ is referred to as {\em observation 
time} and $s$ as {\em waiting time}. One says that the system undergoes
{\em ageing} if $C$ or $R$ depend on both $t$ and $s$ and not merely on the
difference $\tau=t-s$. According to the dynamical scaling alluded to
above, one expects for times $t,s\gg t_{\rm micro}$ and $t-s\gg t_{\rm micro}$,
where $t_{\rm micro}$ is some microscopic time scale, 
the following scaling forms
\BEQ \label{1:gl:SkalCR}
C(t,s) = s^{-b} f_C(t/s) \;\; , \;\; R(t,s) = s^{-1-a} f_R(t/s)
\EEQ
such that the scaling functions $f_{C,R}(y)$ satisfy the following asymptotic
behaviour
\BEQ \label{1:gl:lambdaCR}
f_C(y) \sim y^{-\lambda_C/z} \;\; , \;\; f_R(y) \sim y^{-\lambda_R/z}
\EEQ
as $y\to\infty$ and where $\lambda_C$ and $\lambda_R$, respectively, are known
as the autocorrelation \cite{Fish88,Huse89} and autoresponse exponents
\cite{Pico02}, $a$ and $b$ are further non-equilibrium exponents and
$z$ is the dynamical exponent, where it has been tacitly assumed that the
typical cluster size grows for late times as 
$L(t)\sim t^{1/z}$. The derivation of such growth laws from dynamical scaling
has been studied in great detail \cite{Rute95b}. For a non-conserved 
order parameter and $T<T_c$, the dynamical exponent $z=2$. 
The exponents $\lambda_{C,R}$ are independent of the equilibrium exponents and 
of $z$ \cite{Jans92,Godr02,Cugl02}. 
Since a long time, the equality $\lambda_C=\lambda_R$ had been taken
for granted but recently, examples to the contrary have been found for 
either long-ranged initial correlations in ageing ferromagnets \cite{Pico02} 
or else in the random-phase sine-Gordon (Cardy-Ostlund) model 
\cite{Sche03}. On the other hand, a second-order perturbative analysis
of the time-dependent non-linear Ginzburg-Landau equation reproduces
$\lambda_C=\lambda_R$ \cite{Maze03}. The precise relationship between
$\lambda_C$ and $\lambda_R$ remains to be understood. If one uses an 
infinite-temperature initial state, one has 
$\lambda_C=\lambda_R\geq d/2$ \cite{Yeun96}.  

For ageing ferromagnetic systems with a non-conserved order-parameter, the value
of the exponent $a$ depends on the properties of the equilibrium system as
follows \cite{Henk02a,Henk03e}. A system is said to be in {\em class S} if its 
order-parameter correlator $C_{\rm eq}(\vec{r})\sim \exp(-|\vec{r}|/\xi)$ with
a finite $\xi$ and it is said to be in {\em class L} if  
$C_{\rm eq}(\vec{r})\sim |\vec{r}|^{-(d-2+\eta)}$, where $\eta$ is a standard
equilibrium critical exponent. Then
\BEQ \label{1:gl:a}
a = \left\{ \begin{array}{ll}
1/z          & \mbox{\rm ~~;~ for class S} \\
(d-2+\eta)/z & \mbox{\rm ~~;~ for class L}
\end{array} \right.
\EEQ
For example, in $d>1$ dimensions, the kinetic Ising model with Glauber
dynamics is in class S for temperatures $T<T_c$ and in class L at the
critical temperature $T=T_c$. It is generally accepted that 
$b=0$ for $T<T_c$ and $b=a$ if $T=T_c$, see e.g. \cite{Godr02}. 

The distance from equilibrium is conveniently measured through the
{\em fluctuation-dissipation ratio} \cite{Cugl94a,Cugl94b} 
\BEQ \label{1:gl:FDR}
X(t,s) := T R(t,s) \left( \frac{\partial C(t,s)}{\partial s}\right)^{-1}
\EEQ
At equilibrium, the fluctuation-dissipation theorem states that $X(t,s)=1$. 
Ageing systems may also be characterized through the limit
fluctuation-dissipation ratio
\BEQ
X_{\infty} = \lim_{s\to\infty} \left( \lim_{t\to\infty} X(t,s) \right)
\EEQ
Below criticality, one expects $X_{\infty}=0$, but at $T=T_c$, the value
of $X_{\infty}$ should be universal according to the Godr\`eche-Luck conjecture 
\cite{Godr00a,Godr00b}. This universality has been confirmed in a large 
variety of systems in one and two space dimensions 
\cite{Godr00b,Cala02c,Henk03d,Sast03}. The order of the limits is important, 
since $\lim_{t\to\infty}\left(\lim_{s\to\infty} X(t,s)\right)=1$ always. 

While these statements exhaust the content of dynamical scaling, it may be asked
whether the form of the scaling functions $f_{C,R}(y)$ might be fixed in
a generic, model-independent way through a generalization of that symmetry. 
Indeed, it has been shown \cite{Henk02} that an infinitesimal global 
scale-transformation $t\mapsto (1+\eps)^z t$, $\vec{r} \mapsto (1+\eps)\vec{r}$ 
with a constant $\eps$ such that $|\eps|\ll 1$ can for any given value of $z$ 
be extended to an infinitesimal {\it local scale-transformation} 
where now $\eps=\eps(t,\vec{r})$ may depend on both time and 
space.\footnote{This extension of dynamical scaling has an analogue in critical
equilibrium systems: there global scale invariance can be extended to 
conformal invariance, see \cite{Card96,diFr97,Henk99} for introductions.} 
It can be shown that the local scale-transformations
so constructed act as dynamical symmetries of certain linear field equations
which might be viewed as some effective renormalized equation of motion. 
Practically more important, assuming that the response functions of the
theory transform covariantly under local scale-transformations, the exact
form of the scaling function $f_R(y)$ is found \cite{Henk02,Henk01}
\BEQ \label{1:gl:fR}
f_R(y) = r_0 y^{1+a-\lambda_R/z} \left( y-1 \right)^{-1-a}
\EEQ
where $r_0$ is a normalization constant.\footnote{In order to avoid
misunderstandings, we recall that (\ref{1:gl:fR}) holds for the
total response function as defined in (\ref{1:gl:CR}) {\em without} any
subtractions meant to extract an `ageing part'.} Indeed, in deriving this
result one actually only requests that $R$ transforms covariantly under
under the {\em sub}algebra of the infinitesimal local scale-transformation
which excluded time-translations. We say that a theory where the
$n$-point functions built from certain `quasiprimary' fields transform
covariantly under an algebra of such extended scale-transformations
is {\em locally scale-invariant} \cite{Henk02,Henk01}. 
The prediction (\ref{1:gl:fR}) has been confirmed in a large class of ageing 
ferromagnets as reviewed in \cite{Henk02,Henk03c}. The status of
the scaling function $f_C(y)$ of the spin-spin correlator is less clear, 
however. Building on the Ohta-Jasnow-Kawasaki 
approximation (see \cite{Bray94}) Gaussian closure procedures 
\cite{Bray92,Roja99} in the O($n$)-model produce approximate forms for $f_C(y)$ 
but we do not know of any other approach which does not involve some 
uncontrolled approximation. 

Given the phenomenological success of (\ref{1:gl:fR}), we wish to understand
better where such a supposedly general and exact result derived from a
dynamical symmetry and without using any model-specific properties might come 
from. In this paper, we shall concentrate on the important special case
$z=2$ which describes the phase-ordering kinetics after a 
quench to a temperature $T<T_c$ of a ferromagnet with a 
non-conserved order-parameter \cite{Rute95b}. 
We recall that local scale-transformations are dynamical symmetries
of certain differential equations, such as the free diffusion/free Schr\"odinger
equation for $z=2$. Indeed, the maximal dynamical symmetry of the free diffusion
equation $\partial_t \phi = D \Delta \phi$ is known since a long time to
be the so-called Schr\"odinger group \cite{Lie1882,Goff27,Nied72}, 
to be defined in section~2. 
Also, it is well-established that the same group also describes
the dynamic symmetry of non-relativistic free-field theory \cite{Hage72,Jack72}.
It also arises as dynamical symmetry in certain non-linear Schr\"odinger
equations \cite{Fush87,Ride93,Lahn98,Ghos02,Popo03}, 
the Burgers equation \cite{Katk65,Ivas97} 
or the equations of fluid dynamics \cite{ORai01}. If in addition $D$ is also 
considered as a variable, Schr\"odinger invariance in $d$ dimension becomes
a conformal invariance in $d+2$ dimensions \cite{Henk03}. 
The classification of non-linear
equations and of systems of equations admitting as a dynamical symmetry the
Schr\"odinger group or one of its subgroups (e.g. the Galilei group) 
has received a lot of mathematical attention, 
see \cite{Fush85,Fush89,Fush93,Fush95,Ride93,Cher00,Niki01}. The extension
to dynamical exponents $z\ne 2$ needed for quenches to criticality or for
glassy systems will be left for future work. 

However, the setting just outlined is not yet sufficient for the description
of ageing phenomena. Rather, 
we are interested in the time-dependent behaviour of spin systems coupled
to a heat bath at temperature $T$. It is usually admitted that after 
coarse-graining, this may be modeled in terms of a Langevin equation. 
If there are no macroscopic conservation laws, the Langevin equation for the
coarse-grained order parameter $\phi=\phi(t,\vec{r})$ should be 
model A in the Hohenberg-Halperin classification \cite{Hohe77}
\BEQ \label{2:eq1}
\frac{\partial \phi(t,\vec{r})}{\partial t}=
-D\frac{\delta {\cal H}}{\delta \phi}+\eta(t,\vec{r})
\EEQ
where $\cal H$ is the classical Hamiltonian, and $D$ stands for the diffusion 
constant or equivalently some relaxation rate. Thermal noise is described
by a Gaussian random force $\eta=\eta(t,\vec{r})$ and is
thus characterized by its first two moments 
\BEQ \label{eqeta}
\langle\eta(t,\vec{r})\rangle=0 \;\; , \;\;  
\langle\eta(t,\vec{r})\eta(s,\vec{r}')\rangle=
2DT\,\delta(t-s)\,\delta(\vec{r}-\vec{r}')
\EEQ
where $T$ is the bath temperature. It is well-known \cite{Hohe77,Bray94} that 
this formalism describes the relaxation of the system towards its equilibrium 
state given by  the probability distribution $P_{\rm eq} \sim e^{-{\cal H}/T}$. 
In addition, initial conditions must be taken into account and are described 
in terms of the initial correlation function
\BEQ
a(\vec{r}-\vec{r}') := C(0,0;\vec{r},\vec{r}') := 
\left\langle \phi(0,\vec{r}) \phi(0,\vec{r}')  \right\rangle
\EEQ
and where we already anticipated spatial translation invariance. 

Neither the thermal noise nor the initial correlations
described by $a(\vec{r})$ are included into the local scale-transformations as
studied in \cite{Henk02}
which come from systems such as the free diffusion equation. In this paper,
we want to show how both these sources of fluctuations may be taken into
account and we shall explicitly derive the two-time response and 
correlation functions. Our analysis will be restricted to the case 
where $z=2$ which for instance is already enough to describe ageing below 
criticality. 

As we shall show, it is useful to slightly generalize the problem and
to consider the kinetics of systems which in the simplest case may be 
described by a quadratic Hamiltonian of the form 
\BEQ \label{2:eq2}
{\cal H}[\phi]= \frac{1}{2}\int\!\D t\D\vec{r}\: 
\left[\left(\frac{\partial\phi}{\partial\vec{r}}\right)^{2}+v(t)\phi^{2}\right] 
\EEQ 
where $v(t)$ is a time-dependent external potential. Formally, at the
level of relativistic free-field theory, 
$v(t)$ corresponds to a (time-dependent) mass
squared which would measure the distance from a critical point. Alternatively
$v(t)$ may be viewed as a Lagrange multiplier in order to ensure the
constraint $\langle\phi(t,\vec{r})\phi(t,\vec{r})\rangle=1$ and we shall
make this explicit through the example for the kinetic spherical model in
section~5. In
a physically more appealing way, time-dependent potentials arise when a 
many-body system is brought into contact with a heat bath whose temperature
$T(t)$ is time-dependent \cite{Pico03}. In this paper, we shall be interested
in the dynamics symmetries of Langevin equations derived from a free-field
Hamiltonian (\ref{2:eq2}). In particular, we shall compare the situation without
(i.e. $T=0$) and with thermal noise (i.e. $T>0$). 
For simplicity, we shall refer to all equations
(\ref{2:eq1}) obtained from the Hamiltonian (\ref{2:eq2}) 
as `free Schr\"odinger equations'.

The study of the dynamic symmetries of such free-field theories will yield 
useful insights which we expect to extend to physically more realistic 
interacting field theories where $\cal H$ would also contain higher than merely 
quadratic terms. If we identify $D^{-1}=2\II m$, it is clear  
that the order parameter $\phi$ is given by a noisy Schr\"odinger equation in 
an external time-dependent potential $v(t)$.

This paper is organized as follows. In section 2, we first review the basics of
Schr\"odinger-invariance in the absence of thermal noise and without initial
correlations and then show that through a gauge-transformation
involving $v(t)$, the entire phenomenology of ageing and in  particular
(\ref{1:gl:fR}) can be reproduced. We also consider the selection rules
which follow from Galilei-invariance. In section 3, after having reformulated
the problem in terms of the field-theoretic Martin-Siggia-Rose formalism, 
we study the
effects of thermal noise and/or initial correlations on free-field theory
given by a Hamiltonian (\ref{2:eq2}). In section 4, these results are extended
to any field theory with for $T=0$ and $a(\vec{r})=0$ is Galilei-invariant. 
We find that the two-time response function $R$ is independent of both $T$
and $a(\vec{r})$ and obtain a new reduction formula (\ref{eqC4}) which relates 
$C$ to certain three- and four-point response functions to be evaluated in the
noiseless theory and discuss the scaling of the resulting two-time
autocorrelation function. In sections 5-7, these results are tested
in several exactly solvable systems (with an underlying free-field theory) 
undergoing ageing with $z=2$, 
namely the kinetic spherical model, the XY-model in spin-wave approximation, 
the critical voter model and 
the free random walk. We conclude in section 8. Appendix A deals with technical
aspects of Gaussian integration and appendix~B analyses a special four-point 
response function. In appendix~C we consider a generalized realization of local 
scale-invariance and its application to the $1D$ Glauber-Ising model  

%%%%%%%%%%%%%%%%%%%%%%%%%%%%%%%%%%%%%%%%%%%%%%%%%%%%%%%%%%%%%%%%%%%%%%%%%%%%%%%%
\section{Local scale-invariance: a reminder}
%%%%%%%%%%%%%%%%%%%%%%%%%%%%%%%%%%%%%%%%%%%%%%%%%%%%%%%%%%%%%%%%%%%%%%%%%%%%%%%%

\subsection{Schr\"odinger-invariance}

We begin by reviewing the kinematic symmetries of Schr\"odinger
equations with a time-dependent potential, but without a noise term. 
A long time ago, Niederer \cite{Nied74} obtained the maximal kinematic
symmetry group of the Schr\"odinger equation for an arbitrary
potential $v=v(t,\vec{r})$ 
and he also gave a few examples where that group is isomorphic to the
maximal kinematic group {\sl Sch}$(d)$ of the free Schr\"odinger 
equation \cite{Lie1882,Nied72}. The group {\sl Sch}$(d)$ is called the
{\it Schr\"odinger group} \cite{Nied72}.  

We recall the definition of {\sl Sch}$(d)$. On the time and space coordinates
$(t,\vec{r})$ it acts as $(t,\vec{r})\mapsto (t',\vec{r}')=g(t,\vec{r})$ where 
\BEQ \label{eq1:gl:SCH1}
t \longmapsto t' = \frac{\alpha t + \beta}{\gamma t + \delta} \;\; , \;\;
\vec{r} \longmapsto \vec{r}' = \frac{{\cal R} \vec{r} + \vec{v} t + \vec{a}}
{\gamma t + \delta} \;\; ; \;\; \alpha\delta - \beta\gamma =1 
\EEQ
where $\cal R$ is a rotation matrix. 
The action of {\sl Sch}$(d)$ on the space of solutions
$\phi$ of the free Schr\"odinger equation is projective, that is,
the wave function $\phi=\phi(t,\vec{r})$ 
transforms into
\BEQ \label{4:phitrans}
\phi(t,\vec{r}) \longmapsto \left( T_g \phi\right)(t,\vec{r}) =
f_g[g^{-1}(t,\vec{r})] \phi[g^{-1}(t,\vec{r})]
\EEQ
and the companion function $f_g$ is explicitly known \cite{Nied72,Nied74}. 
The projective unitary irreducible representations of {\sl Sch}$(d)$ are
classified \cite{Perr77}. We now carry this over to field theory and consider
fields transforming according to (\ref{4:phitrans}). By analogy with an
analogous terminology in conformal field theory \cite{Bela84}, 
a field $\phi$ transforming according
to (\ref{4:phitrans}) and with $g(t,\vec{r})$ given by (\ref{eq1:gl:SCH1}) is
called {\em quasiprimary} \cite{Henk02}. Schr\"odinger-invariance is the
$z=2$ special case of local scale-invariance, with time-translations added. 

Besides the examples given in \cite{Nied74}, there exist
further noiseless Schr\"odinger equations with a maximal kinematic group  
isomorphic to {\sl Sch}$(d)$. Consider the noiseless Langevin 
equation  
\BEQ 
D^{-1} \frac{\partial \phi(t,\vec{r})}{\partial t}
= \Delta\phi(t,\vec{r}) - v(t)\phi(t,\vec{r}) 
\label{eqSPE}
\EEQ
where $D$ is the diffusion constant. This equation can be reduced to the  
free Schr\"odinger equation 
$D^{-1}\partial_t \Psi(t,\vec{r})=\Delta\Psi(t,\vec{r})$ through the 
gauge transformation 
\BEQ \label{eqgauge}
\phi(t,\vec{r})=\Psi(t,\vec{r})
\exp\left(-D\int_{0}^{t}\!\D u\, {v(u)}\right) 
\EEQ 
Since the kinematic symmetries of the free Schr\"odinger equation are 
well-understood and the realization of the Schr\"odinger group is
explicitly known, the corresponding realization for the case at hand,
similar to (\ref{4:phitrans}), readily follows.

It turns out that the only change occurs in the companion function $f_g$. 
Let $f_{g}^{(0)}$ stand for the companion function of the free
Schr\"odinger equation, then because of the gauge transformation 
eq.~(\ref{eqgauge}) we find
\BEQ \label{eqfg}
f_g(t,\vec{r})=f_g^{(0)}(t',\vec{r})\exp\left(-D\int_t^{t'}\!\D u\,v(u)\right)
\EEQ
where $t'=t'(t)$ has been defined in eq.~(\ref{eq1:gl:SCH1}). 
The generators of the Lie algebra $\mathfrak{sch}_1$ of this realization  
of {\sl Sch}$(1)$, appropriate
for the equation (\ref{eqSPE}) with the potential $v(t)$, read 
\BEA 
X_{-1} = -\partial_t+\frac{v(t)}{2\cal M} & & 
\mbox{\rm time drift} \nonumber \\ 
X_{0} = -t\partial_t -\frac{1}{2}r\partial_r-\frac{x}{2}+t\frac{v(t)}{2\cal M}
& &
\mbox{\rm dilatation} \nonumber \\
X_{1} = -t^2\partial_t - tr\partial_r -xt + t^{2}\frac{v(t)}{2\cal M}
- \frac{\cal M}{2} r^2& &
\mbox{\rm special Schr\"odinger transformation} \nonumber \\
Y_{-1/2} = -\partial_r & &
\mbox{\rm space translation} \nonumber \\
Y_{1/2} = - t\partial_r - {\cal M} r & &
\mbox{\rm Galilei transformation} \nonumber \\
M_{0} = - {\cal M} & &
\mbox{\rm phase shift}.
\label{eqgenSPE}
\EEA
where we expressed the diffusion constant $D^{-1}=2{\cal M}$ as the `mass'
$\cal M$ and $x$ denotes the scaling dimension of the wave function 
$\phi(t,\vec{r})$. Of course, $x=d/2$ for a solution of the
free Schr\"odinger equation, but it will be useful to consider arbitrary values
of $x$ as well.     

The non-vanishing commutators of the Lie algebra $\mathfrak{sch}_1$ spanned by 
the generators (\ref{eqgenSPE}) are
\BEQ \label{2:gl:SCHcomm}
\left[ X_n, X_{n'} \right] = (n-n') X_{n+n'} \;\; , \;\;
\left[ X_n, Y_m \right] = \left(\frac{n}{2}-m\right) Y_{n+m} \;\; , \;\;
\left[ Y_{1/2}, Y_{-1/2}\right] = M_{0} 
\EEQ 
where $n,n'\in \{\pm 1,0\}$, $m\in\{\pm\frac{1}{2}\}$ and  
with straightforward extensions to $d>1$, see \cite{Henk02}.

\subsection{Galilei-covariance of correlators}

When discussing the dynamic symmetries of a time-dependent statistical system,
the requirement of Galilei-invariance plays a particular r\^ole. Indeed, for
a system with local interactions and which is invariant under
space translations, scale transformations with a dynamical exponent $z=2$ and
in addition Galilei-invariant, it can be shown that there exists a Ward identity
such that the system is also invariant under the `special' Schr\"odinger
transformation (\ref{eqgenSPE}) \cite{Hage72,Henk94,Henk03}. 

We shall be particularly interested in the two-point 
correlator $C$ and the linear response function $R$ built from the
order parameter $\phi(t,\vec{r})$. Using Martin-Siggia-Rose theory (MSR theory)
which we shall briefly review in section~3, these may be expressed in
terms of $\phi$ and the so-called response field $\wit{\phi}$ as follows
\BEA 
C(t,s;\vec{r},\vec{r}') &:=& \langle \phi(t,\vec{r})\phi(s,\vec{r}') \rangle 
\nonumber \\
R\left( t,s;\vec{r},\vec{r}' \right) &:=& \left.\frac{\delta 
\langle \phi(t,\vec{r}) \rangle}{\delta h(s,\vec{r}')}\right|_{h=0}=
\langle \phi(t,\vec{r}){\wit \phi}(s,\vec{r}') \rangle 
\label{3:gl:CRC} \\
\wit{C}\left( t,s;\vec{r},\vec{r}' \right) &:=&
\langle \wit{\phi}(t,\vec{r})\wit{\phi}(s,\vec{r}') \rangle
\nonumber 
\EEA
where $h$ is the magnetic field conjugate to the order parameter $\phi$.  
Later, we shall often refer to $t$ as the {\it observation time} and to $s$ 
as the {\it waiting time}. 

Generalizing the above definition to $v(t)\ne 0$, we say that a field
$\phi$ is {\em quasiprimary} if its infinitesimal change under 
$\mathfrak{sch}_1$ is given by the generators (\ref{eqgenSPE}) with 
$v(t)\ne 0$. A quasiprimary field $\phi$ is characterized by its scaling 
dimension $x$ and its `mass' ${\cal M}\geq 0$. 
In turn, if the response field $\wit{\phi}$ associated to $\phi$ is also
quasiprimary, it has a scaling dimension denoted by $\wit{x}$ and the `mass' 
\BEQ \label{2:gl:Mtilde}
\wit{\cal M}=-{\cal M} \leq 0
\EEQ
This important fact will be used later on. 
The argument leading to the result (\ref{2:gl:Mtilde}) was discussed in detail 
in \cite{Henk03} and will not be repeated here. 

If both $\phi$ and $\wit{\phi}$ transform
as quasiprimary fields of a Schr\"odinger-invariant theory, 
the generators eqs.~(\ref{eqgenSPE}) can be used 
to derive restrictions on the form of any multipoint correlator 
and in particular determine the two-point functions completely. 
If ${\cal X}_i$ is any of the generators of $\mathfrak{sch}_d$ acting
on the $i^{\rm th}$ particle in a $n$-point correlator 
$\mathfrak{A}\{t_i,\vec{r}_i\}$ where $i=1,\ldots,n$ 
(see (\ref{3:gl:CRC}) for $n=2$), 
we have a set of differential equations
\BEQ
\left( {\cal X}_1 + \ldots + {\cal X}_n \right) 
\mathfrak{A}\{t_i,\vec{r}_i\} = 0
\EEQ

If rotation invariance can be assumed (and we shall implicitly do so throughout
this paper), for the calculation of the two-point
functions it is enough to consider the one-dimensional case and use the
generators of eq.~(\ref{eqgenSPE}). Then a straightforward calculation 
\cite{Poly70} gives, provided the `mass' of the order parameter is positive 
${\cal M}>0$, see e.g. \cite{Henk94,Henk02}
\BEA
\wit{C}_0 \left( t,s;\vec{r},\vec{r}' \right)=0 \nonumber \\
C_0\left( t,s;\vec{r},\vec{r}' \right)=0 
\label{eqcor}
\EEA 
whereas the response function is basically the gauge-transformed expression 
of the well-known zero-potential Gaussian response $\cal R$, i.e. 
\BEA
R_0\left( t,s;\vec{r},\vec{r}' \right) &=&
\frac{k(t)}{k(s)}{\cal R}\left( t,s;\vec{r},\vec{r}' \right) 
\nonumber \\
{\cal R}\left( t,s;\vec{r},\vec{r}' \right) &=&
\delta_{x,\wit{x}}\: r_{0}\,\Theta(t-s) \left( t-s \right)^{-x}
\exp \left( -\frac{\cal M}{2}\frac{(\vec{r}-\vec{r}')^{2}}{(t-s)} \right)
\nonumber \\
k(t) &:=& \exp\left(-\frac{1}{2{\cal M}}\int^{t}\,\D u\: v(u)\right)
\label{eqres}
\EEA
where $r_0$ is a normalization constant and $\Theta$ is the 
Heaviside function which expresses causality. As they stand,
eqs.~(\ref{eqcor},\ref{eqres}) hold for $T=0$ and we shall from now on use
the index $0$ to remind the reader of this fact. 

On the other hand, if the system is not rotation-invariant, 
we can repeat the same argument in any fixed direction of space and 
the non-universal constant $\cal M$ 
becomes direction-dependent. Indeed, rotation invariance is broken
for phase-ordering systems defined on a lattice \cite{Rute96,Rute99} for
sufficiently small temperatures. Even then, local scale invariance still
holds in every single space direction, as exemplified in the $2D$ and 
$3D$ Ising models with Glauber dynamics \cite{Henk03b}.  

A few comments are in order.
\begin{enumerate}
\item 
Eqs.~(\ref{eqcor},\ref{eqres}) provide a manifest example of the 
superselection rule of Galilei invariance, also known as Bargman 
superselection rule \cite{Barg54}. Explicitly, if $\Phi_i(t_i,\vec{r}_i)$ 
are Galilei-covariant fields, each with a `mass' ${\cal M}_i$, 
Galilei-covariance implies \cite{Barg54,Henk94}
\BEQ \label{eqgal}
\left\langle \Phi_1(t_1,\vec{r}_1)\ldots \Phi_n(t_n,\vec{r}_n)
\right\rangle = \delta_{{\cal M}_1+\ldots+{\cal M}_n,0}\: 
F\left( \left\{ t_i, \vec{r}_i\right\}\right)
\EEQ
By physical convention, the `masses' of the fields $\phi$ are 
non-negative, viz. ${\cal M}_i\geq 0$. Furthermore, the response fields 
$\wit{\phi}$ have {\em negative}  `masses' $\wit{\cal M}_i\leq 0$ and the
result (\ref{eqcor}) follows. 
\item 
In the introduction, we reviewed the result (\ref{1:gl:fR}) for the 
autoresponse function, derived for arbitrary $z$ from a generalization of
Schr\"odinger invariance \cite{Henk02,Henk01}. 
For $z=2$, eq.~(\ref{1:gl:fR}) co\"{\i}ncides with 
our result (\ref{eqres}) provided that
\BEA
x &=& {\wit x} \; = \; 1+a
\nonumber \\
v(t) &=& \left(2{\cal M}\right)\,\frac{1+a-\lambda_{R}/2}{t}
\label{eqsc}
\EEA
\item 
However, there is an important difference in our derivation of the scaling
form of $R_0(t,s)$ with respect to \cite{Henk02,Henk01}: because
time-translation invariance is broken in ageing phenomena, covariance of
$R_0$ was required to hold merely under a subalgebra of $\mathfrak{sch}_d$ 
where the time-translations were excluded. Indeed, in \cite{Henk03}, such
subalgebras were studied systematically and a relationship with the 
parabolic subalgebras of the (complexified) conformal algebra
$\mathfrak{conf}_{d+2}$ was found. On the other hand, the realization of
$\mathfrak{sch}_d$ used in \cite{Henk02,Henk01,Henk03} applies to a vanishing 
potential $v(t)=0$. 

Here, we do not follow that point of view. We consider the more general
realization of the entire algebra $\mathfrak{sch}_1$ with a time-dependent 
potential $v(t)$ and require that both $\phi$ and ${\wit \phi}$ transform as 
quasiprimary fields under the {\it whole} set of generators (\ref{eqgenSPE}).

That these two different approaches, given the conditions (\ref{eqsc}), yield 
the {\em same} phenomenology of the scaling of the two-time autoresponse 
function is our first result and will be crucial for the developments to follow.
\item
At first sight, the result (\ref{eqcor}) that the two-time autocorrelator
$C(t,s)=\langle \phi(t,\vec{r})\phi(s,\vec{r})\rangle = 0$ may appear strange
and indeed does not hold true in concrete models. In the next
sections, we shall show that this apparent contradiction comes from the
fact that the noiseless Schr\"odinger equation does not take the thermal noise
into account. As we shall see, the reformulation of Schr\"odinger invariance
in ageing systems in terms of a noisy Schr\"odinger equation with a
time-dependent potential allows to arrive at physically meaningful predictions
for correlation functions. Explicit confirmations in exactly soluble models
will be presented.   
\end{enumerate}

%%%%%%%%%%%%%%%%%%%%%%%%%%%%%%%%%%%%%%%%%%%%%%%%%%%%%%%%%%%%%%%%%%%%%%%%%%%%%%%%
\section{Response and correlation functions for non-interacting Gaussian 
theories}
%%%%%%%%%%%%%%%%%%%%%%%%%%%%%%%%%%%%%%%%%%%%%%%%%%%%%%%%%%%%%%%%%%%%%%%%%%%%%%%%

\subsection{The Martin-Siggia-Rose formalism}

It is useful to treat noisy Langevin equations in the context of the
Martin-Siggia-Rose (MSR) formalism \cite{Mart73,Jans92,Zinn89,Cugl02,Taeu04}. 
In equilibrium, the integration over the Gaussian noise $\eta$ can be 
carried out
by introducing a response field $\wit{\phi}$.\footnote{In the systematic
terminology of \cite{Card96}, this should be rather called a {\em response
operator}, because $\wit{\phi}$ will become an operator in a canonical
quantization scheme of the action. The notion of a response field should
have been reserved to the canonically conjugate variable of $\wit{\phi}$. 
Since we shall not use the operator formalism here, we shall simply, but
sloppily, talk about $\wit{\phi}$ as a response field.} It can be shown 
that the stochastic Langevin equation (\ref{2:eq1}) can be obtained from the
following effective action $\Sigma[\phi,\wit{\phi}]$ 
\BEQ
\Sigma[\phi,\wit{\phi}]= \int \!\D t\,\D\vec{r}\: 
\wit{\phi}\left(\frac{\partial \phi}{\partial t}
+D\frac{\delta{\cal H}}{\delta \phi}\right)
-\frac{1}{2}\int\!\D t\,\D\vec{r}\,\D t'\,\D\vec{r}'\:
\langle\eta(t,\vec{r})\eta(t',\vec{r}')\rangle\:
\wit{\phi}(t,\vec{r}) \wit{\phi}(t',\vec{r}') 
\label{eqMSRgen}
\EEQ
This action appears in the generating functional 
$Z=\int\!{\cal D}\phi{\cal D}\wit{\phi}\:e^{-\Sigma[\phi,\wit{\phi}]}$
expressed as a path integral. In this way, the original dynamical problem in
$d$ dimensions has been mapped onto one of statistical mechanics in
$d+1$ dimensions. 

As long as one is merely interested in equilibrium behaviour and provided the
dynamics is ergodic, there is no need to worry about initial conditions, 
which might be said to be specified
at a time $t=-\infty$. Here, we are interested in how 
the equilibrium state is reached from a given initial state and must
include into the action a term describing  
the initial preparation of the system. One has
\BEQ  
S[\phi,\wit{\phi}] = \Sigma[\phi,\wit{\phi}]  + \sigma[\phi,\wit{\phi}]
\label{eqdifact}
\EEQ
where $\Sigma[\phi,\wit{\phi}]$ (\ref{eqMSRgen}) describes the `bulk' evolution 
of the system as derived from the Langevin equation while
$\sigma[\phi,\wit{\phi}]$ describes the initial conditions at time $t=0$.
As already pointed out by Mazenko \cite{Maze03}, it may be written as
\BEQ
\sigma[\phi,\wit{\phi}]=-\frac{1}{2}
\int_{\mathbb{R}^{d}\times\mathbb{R}^{d}} \!\D\vec{r}\,\D\vec{r}'\:  
{\wit{\phi}}(0,\vec{r})a(\vec{r}-\vec{r}'){\wit{\phi}}(0,\vec{r}')
\label{eqboun2}
\EEQ
where it is implicitly admitted that $\langle \phi(0,\vec{r})\rangle =0$ and
$a(\vec{r})$ is the initial two-point correlator 
\BEQ
a(\vec{r}) := C(0,0;\vec{r}+\vec{r}',\vec{r}') = 
\langle \phi(0,\vec{r}+\vec{r}') \phi(0,\vec{r}')\rangle
\EEQ
{}From spatial translation-invariance, it follows that $a(\vec{r})=a(-\vec{r})$ 
which we shall admit throughout.
%%MH For the sake of simplicity, we admit a parity-symmetric initial
%%MH correlator $a(\vec{r})=a(-\vec{r})$ throughout. 

We call the theory described by the action $S_0$ alone the
{\em noiseless} theory. 

For a free field, the noiseless and the thermal parts of the MSR action 
read (we also have set $D=1$)
\BEA
S_{0}[\phi, {\wit \phi}] &:=& \int \!\D t\,\D\vec{r}\: 
\wit{\phi}\left(\frac{\partial \phi}{\partial t}-\Delta\phi+v(t)\phi\right)
\nonumber \\
{\cal S}[\phi, {\wit \phi}] &:=& -T\int \!\D t\,\D\vec{r}\: 
{\wit{\phi}}^{2}(t,\vec{r}) 
\nonumber \\
\Sigma[\phi, {\wit \phi}] &=& S_{0}[\phi,{\wit \phi}]+
{\cal S}[\phi, {\wit \phi}]
\label{eqbulktherm}
\EEA
We shall refer to the contribution described by $\cal S$
and $\sigma$ as the {\em thermal} and {\em initial} noise, respectively. 

We point out that field-theoretic studies of critical dynamics use a
different initial term, namely \cite{Jans89}
\BEQ \label{eqboun}
\sigma_c[\phi,\wit{\phi}]=\frac{\tau_{0}}{2} \int_{\mathbb{R}^d} \!\D\vec{r}\: 
\left(\phi(0,\vec{r})-m(\vec{r})\right)^{2}
\EEQ
This specifies an initial macroscopic state with spatially varying 
order parameter $\left\langle\phi(0,\vec{r})\right\rangle=m(\vec{r})$ 
and spatial correlations decaying on a finite scale 
proportional to ${\tau_{0}}^{-1}$. Ageing {\it at} criticality
was studied in the O($n$)-model using the $\eps$-expansion
with the initial term $\sigma_c$ \cite{Jans89,Cala02a,Cala02b,Cala02c,Cala03}. 
However, the use of $\sigma_c$ instead of $\sigma$ for temperatures below
criticality would lead to contradictions. 

Treating the noisy Langevin equation (\ref{2:eq1}) as the classical equation
of motion of the field-theory (MSR) action 
eqs.~(\ref{eqdifact},\ref{eqbulktherm},\ref{eqboun2}) has the following
advantages. 
\begin{itemize}
\item Thermal fluctuations and initial conditions are explicitly included. 
To emphasize this, notice that the noisy contributions to the 
MSR action are
\BEA
S[\phi, {\wit \phi}]-S_{0}[\phi, {\wit \phi}] &=& -
\int \!\D u\,\D \vec{r}\,\D u'\,\D\vec{r}'\:  
{\wit \phi}(u,\vec{r})\kappa(u,u';\vec{r}-\vec{r}') {\wit \phi}(u',\vec{r}')
\nonumber \\
\kappa(u,u';\vec{r}) &=& T\delta(u-u')\delta(\vec{r}) 
+\frac{1}{2}\delta(u)\delta(u')a(\vec{r})
\label{equni}
\EEA     
where $\kappa$ includes both the effects of thermal and of initial-state 
fluctuations.                        

\item The response field ${\wit \phi}$ describes the thermal noise as can be
seen from the equations of motion derived from the free-field 
MSR action (\ref{eqbulktherm})
\BEA
\frac{\partial \phi(t,\vec{r})}{\partial t} &=&
\Delta\phi(t,\vec{r})-v(t)\phi(t,\vec{r})+2T\wit{\phi}(t,\vec{r})
\nonumber \\
-\frac{\partial \wit{\phi}(t,\vec{r})}{\partial t} &=& 
\Delta\wit{\phi}(t,\vec{r})-v(t)\wit{\phi}(t,\vec{r})
\label{5:eq:motion}
\EEA 
The first equation (\ref{5:eq:motion}) reduces to the Langevin equation 
(\ref{2:eq1})  provided one makes the formal identification 
\BEQ
\eta(t,\vec{r})=2T\wit{\phi}(t,\vec{r})
\label{eqnoires}
\EEQ 
Therefore, at the classical level, the stochastic Langevin 
equation eq.~(\ref{2:eq1}) is described by two deterministic equations. 
In our case they are both of Schr\"odinger type and with opposite masses for 
the field $\phi$ and the response field ${\wit \phi}$.           

%%The second equation (\ref{5:eq:motion}) depends only on the response field, 
%%and is a Schr\"odinger equation with a time-dependent potential and a
%%mass with the opposite sign with respect to the one of $\phi$. 

\item Averages of any $n$-point function built from the fields 
$\phi, \wit{\phi}$ can be expressed
in terms of the functional integral 
\BEQ
\left\langle F\{\phi(t_i,\vec{r}_i),\wit{\phi}(t_j,\vec{r}_j)\}\right\rangle=
\int\!{\cal D}\phi\,{\cal D}\wit{\phi}\: 
\exp\left(-S[\phi,\wit{\phi}]\right) 
F\{\phi(t_i,\vec{r}_i),\wit{\phi}(t_j,\vec{r}_j)\} 
\label{eqnpoints}
\EEQ
with the normalization $\langle 1 \rangle =1$. 

For example adding a magnetic perturbation
$\delta{\cal H}_{\rm mag}=-h\phi$ to the Hamiltonian (\ref{2:eq2}) and 
then computing the mean of the order-parameter to first order in 
$h$, the relation 
$R(t,s;\vec{r},\vec{r}')=\langle\phi(t;\vec{r}){\wit \phi}(s,\vec{r}')\rangle$ 
of eq.~(\ref{3:gl:CRC}) is easily reproduced. 

\item The use of Martin-Siggia-Rose formalism makes the machinery of the 
field-theoretic renormalization-group available \cite{Zinn89,Card96,Taeu04} 
but we shall not pursue this here. 
\end{itemize}
It will be useful to split the calculation of averages into two steps as
follows
\BEA
\left\langle F\{\phi(t_i,\vec{r}_i),\wit{\phi}(t_j,\vec{r}_j)\}\right\rangle 
&=& \left\langle F\{\phi(t_i,\vec{r}_i),\wit{\phi}(t_j,\vec{r}_j)\}
\exp\left(-{\cal S}[\phi, {\wit \phi}]
-\sigma[\phi, {\wit \phi}]\right)\right\rangle_{0}
\nonumber \\
\left\langle F\{\phi(t_i,\vec{r}_i),\wit{\phi}(t_j,\vec{r}_j)\}\right\rangle_0 
&=& \int\!{\cal D}\phi\,{\cal D}\wit{\phi}\: 
\exp\left(-S_{0}[\phi,\wit{\phi}]\right) 
F\{\phi(t_i,\vec{r}_i),\wit{\phi}(t_j,\vec{r}_j)\} 
\label{eqnpoints2}
\EEA     
where the notation $\langle \rangle_{0}$ (and more generally the index $0$) 
refers from now on to averages of the non-fluctuating theory. 
This allows to make use of the Schr\"odinger invariance of the noiseless
theory.     

\subsection{Analytical results for free fields}
In this section, we find both response and correlation 
functions for free-field Martin-Siggia-Rose theory, as given by
eqs.~(\ref{eqdifact},\ref{eqboun2},\ref{eqbulktherm}).  

\subsubsection{Two-point functions without noise}
 
The free Martin-Siggia-Rose action  
$S[\phi, {\wit \phi}]$ has a Gaussian structure. We shall write it as
\BEQ \label{3:gl:S}
S[\phi, {\wit \phi}]= \int \!\D u\,\D \vec{r}\,\D u'\,\D\vec{r}'\: 
\Phi(u;\vec{r})^T{\cal Q}(u,u';\vec{r},\vec{r}') \Phi(u';\vec{r}')
\EEQ   
where $\Phi=\left(\vekz{\phi}{\wit{\phi}}\right)$ is the two-component 
field built from $\phi$ and ${\wit \phi}$, and $\Phi^{T}$ stands for its 
transpose. The kernel ${\cal Q}$ reads
\BEQ
{\cal Q}(u,u';\vec{r},\vec{r}')=\frac{1}{2}\left[
\begin{array}{cc}
0 & \delta(u-u')\delta(\vec{r}-\vec{r}')\left(-\Delta-\partial_u\right) 
+\frac{1}{2} v(u)\\
\delta(u-u')\delta(\vec{r}-\vec{r}')\left(-\Delta+\partial_u\right)
+\frac{1}{2} v(u) & 
-2\kappa(u,u';\vec{r}-\vec{r}')
\end{array}
\right]
\EEQ 
This peculiar form of the Lagrangian density is quite suggestive as regards the 
Galilei-invariance. When $\kappa=0$, that is in absence of noise, the 
quadratic form $\cal Q$ is antidiagonal. This is one way of presenting the
Bargman superselection rules and leads in particular to
$\langle\phi\phi\rangle =\langle\wit{\phi}\wit{\phi}\rangle =0$ which is
a manifestation of Galilei-invariance of the noiseless system.  
The presence of noise just breaks this symmetry.

In order to study systematically the r\^ole of the noise, 
we shall expand around the non-fluctuating theory. The correlation functions 
$C_{0}, {\wit C_{0}}$ and the linear response function $R_{0}$ are
\BEA
C_{0}(t,s;\vec{r},\vec{r}') &=& 0
\nonumber \\
{\wit C}_{0}(t,s;\vec{r},\vec{r}') &=& 0
\nonumber \\
R_0\left( t,s;\vec{r},\vec{r}' \right) &=&
\frac{k(t)}{k(s)}\Theta(t-s) \left( 4\pi(t-s) \right)^{-d/2}
\exp \left( -\frac{1}{4}\frac{(\vec{r}-\vec{r}')^{2}}{(t-s)} \right)
\label{eq0MSR}
\EEA
This follows since the 
quadratic form ${\cal Q}$ is antidiagonal the field $\phi$ can only be coupled 
to ${\wit \phi}$ and the result is just the bare propagator of the 
theory.  It is clear that only the antidiagonality of $\cal Q$ is important to 
derive this result and the explicit free-field form (\ref{3:gl:S}) is
not required for (\ref{eq0MSR}) to hold. 

These results fully agree with the Schr\"odinger-invariance prediction
eq.~(\ref{eqres},\ref{eqcor}) with the identifications $x={\wit x}=d/2$, 
${\cal M}=1/2$ and $r_{0}=(4\pi)^{-d/2}$.

A further manifestation of the Galilei invariance of the noiseless theory is
the fact that 
\BEQ \label{3:gl:Barg}
\langle\,\underbrace{\phi\cdots\phi}_n \;
\underbrace{\wit{\phi}\cdots\wit{\phi}}_m\,\rangle_0=0
\EEQ
unless $n=m$ as is easily checked. This is a further example of the
Bargman superselection rule (\ref{eqgal}) 
and will be important in what follows.

\subsubsection{Two-point functions in presence of noise}

We now find the same two-point functions in the presence 
of noise. 

We begin with the response functions $R(t,s;\vec{r};\vec{r}')$ which is 
found from (\ref{eq0MSR}) by averaging with the noiseless weight 
$\exp\left(-S_{0}[\phi, {\wit \phi}]\right)$  
\BEQ
R(t,s;\vec{r},\vec{r}')=\left\langle  
\phi(t,\vec{r})\wit{\phi}(s,\vec{r}')\exp\left( 
\int \!\D u\,\D \vec{R}\,\D u'\,\D\vec{R}'\: 
{\wit \phi}(u,\vec{R})\kappa(u,u';\vec{R}-\vec{R}') 
{\wit \phi}(u',\vec{r}')\right) \right\rangle_{0} 
\label{eqR1}
\EEQ 
where $\kappa$ is given by eq.~(\ref{equni}). 
Formally expanding the exponential and taking the Bargman superselection rule
(\ref{3:gl:Barg}) into account, only the term of lowest order remains. 
We thus find
\BEQ
R(t,s;\vec{r},\vec{r}')=R_{0}(t,s;\vec{r},\vec{r}')=
\frac{k(t)}{k(s)}\Theta(t-s) \left( 4\pi(t-s) \right)^{-d/2}
\exp \left( -\frac{1}{4}\frac{(\vec{r}-\vec{r}')^{2}}{(t-s)} \right)
\label{eqR2}
\EEQ    
and we see that $R$ is independent of the noise. 

Next, the order-parameter correlation function reads
\BEQ
C(t,s;\vec{r},\vec{r}')=\left\langle 
\phi(t,\vec{r})\phi(s,\vec{r}')\exp\left( 
\int \!\D u\,\D \vec{R}\,\D u'\,\D\vec{R}'\: {\wit \phi}(u,\vec{R})
\kappa(u,u';\vec{R}-\vec{R}') {\wit \phi}(u',\vec{r}')\right) \right\rangle_{0} 
\label{eqC1}
\EEQ
Again expanding the exponential and using eq.~(\ref{3:gl:Barg}), a single
term remains and we readily find
\BEA
C(t,s;\vec{r},\vec{r}') &=& \int\!\D\vec{R}\,\D u\,\D\vec{R}'\,\D u'\: 
\kappa(u,u';\vec{R}-\vec{R}')
R_{0}^{(4)}(t,s,u,u';\vec{r},\vec{r'},\vec{R},\vec{R}')
\nonumber \\
R_{0}^{(4)}(t,s,u,u';\vec{r},\vec{r'},\vec{R},\vec{R}') &:=& \left\langle 
\phi(t,\vec{r})\phi(s,\vec{r}'){\wit \phi}
(u,\vec{R}){\wit \phi}(u',\vec{R}')\right\rangle_{0}
\label{eqC2}
\EEA  
where $R_{0}^{(4)}$ is a noiseless four-point function. 

Repeating the same arguments as before, it is an easy task to compute the 
two-point correlations of response fields. They are
\BEQ \label{eqCt1}
{\wit C}(t,s;\vec{r},\vec{r}')=0
\EEQ
We emphasize that eqs.~(\ref{eqR1},\ref{eqC2},\ref{eqCt1}) will hold for any 
theory satisfying the Bargman's superselection rule (\ref{3:gl:Barg}). We shall
come back to this in section~4. 

The four-point function $R_{0}^{(4)}$ is simple to access for free
fields since it factorizes into a product of two-point functions because
of Wick's theorem. We have
\BEQ \label{eqWick}
R_{0}^{(4)}(t,s,u,u';\vec{r},\vec{r'},\vec{R},\vec{R}')=
R_{0}(t,u;\vec{r},\vec{R})R_{0}(s,u';\vec{r'},\vec{R}')
+R_{0}(t,u';\vec{r},\vec{R}')R_{0}(s,u;\vec{r'},\vec{R})
\EEQ  
Together with the explicit form of $\kappa$, this yields the final result  
\BEA
C(t,s;\vec{r},\vec{r}') &=& C_{th}(t,s;\vec{r},\vec{r}')
+C_{pr}(t,s;\vec{r},\vec{r}')
\nonumber \\
C_{th}(t,s;\vec{r},\vec{r}') &=&  
2T\int\!\D u\,\D\vec{y}\: 
R_{0}(t,u;\vec{r},\vec{y}) R_{0}(s,u;\vec{r}',\vec{y})\, 
\nonumber \\
C_{pr}(t,s;\vec{r},\vec{r}') &=& \int 
\!\D\vec{y}\,\D\vec{y}'\: R_{0}(t,0;\vec{r},\vec{y})
a(\vec{y}-\vec{y}')R_{0}(s,0;\vec{r}',\vec{y}')
\label{eqcornoise}
\EEA
where we separated $C$ in a thermal term $C_{th}$ and an initial term $C_{pr}$. 
We clearly see that while the only contributions to $C$ come from the noise,
$R$ does not depend on it. 

We summarize the results obtained so far as follows:
\begin{enumerate}
\item It is satisfying that the well-known result 
$\wit{C}=\langle\wit{\phi}\wit{\phi}\rangle=0$ is naturally reproduced, see
(\ref{eqCt1}). 
\item The {\em independence} eq.~(\ref{eqR2}) 
of the two-time response function of $T$ and of the initial correlations goes
beyond the usual scaling arguments as reviewed in section 1. 
This explains to some extent the success of the existing confirmations
of that prediction of local scale invariance. 
\item We arrive at an explicit expression for the
two-time correlators, which are obtained in terms of a 
contraction of two response functions. We also see that the earlier result
$C=0$ comes from neglecting both initial and thermal fluctuations. 
\end{enumerate} 

It is useful to present these results also in momentum space. Using spatial
translation invariance, define the Fourier transform of any two-point function 
$A(t,s;\vec{x},\vec{y})=A(t,s;\vec{x}-\vec{y})$ as 
\BEQ
\wht{A}(t,s;\vec{q}) := \int_{\mathbb{R}^d} \!\D\vec{r}\:A(t,s;\vec{r})
\exp\left(-\II\vec{q}\cdot\vec{r}\right)
\label{eqfourier}
\EEQ
Then eqs.~(\ref{eqR2},\ref{eqcornoise}) become
\BEA
\wht{R}(t,s;\vec{q}) &=& \frac{k(t)}{k(s)}\exp\left(-{\vec q}^{2}(t-s)\right)
\,\Theta(t-s) \nonumber \\
\wht{C}_{th}(t,s;\vec{q}) &=&
2T\int_{0}^{s} \!\D u\:\frac{k(t)k(s)}{k^{2}(u)} 
\exp \left(-{\vec{q}}^{2}(t+s-2u) \right)  
\nonumber \\
\wht{C}_{pr}(t,s;\vec{q}) &=&
\wht{R}_{0}(t,0;\vec{q})
\wht{a}(\vec{q})\wht{R}_{0}(s,0;-\vec{q})
=\wht{a}(\vec{q})\frac{k(t)k(s)}{k^{2}(0)}
\exp\left(-{\vec{q}}^{2}(t+s)\right)  
\label{eqobsfour}
\EEA 
where in the second line, the convention $s<t$ has been used. 
In particular, if $v(t)=0$ the two-point 
correlation function takes an especially simple form 
$\wht{C}^{0}(t,s;\vec{q})$  where
\BEQ
\wht{C}^{0}(t,s;\vec{q})=\left[ \wht{a}(\vec{q})-\frac{T}{{\vec{q}}^{2}} \right]
\exp \left(-{\vec{q}}^{2}(t+s)\right) +  
\frac{T}{{\vec{q}}^{2}}\exp\left(-{\vec{q}}^{2}(t-s)\right)
\label{eqcorfou}
\EEQ  
While both response and correlation functions depend on both $t$ and $s$
and therefore describe an ageing behaviour, there is an 
{\it equilibrium regime} $1\ll t-s\ll s,t$ where we have the
following simple expressions 
\BEA
\wht{R}_{\rm eq}(t,s;\vec{q}) = \exp\left(-{\vec q}^{2}(t-s)\right)
\nonumber \\
\wht{C}^{0}_{\rm eq}(t,s;\vec{q}) = 
\frac{T}{{\vec{q}}^{2}}\exp\left(-{\vec{q}}^{2}(t-s)\right)
\label{eqeq}
\EEA
More generally, it is not difficult to show that 
$\wht{C}_{\rm eq}(t,s;\vec{q})=\left( T/\vec{q}^2\right) e^{-\vec{q}^2(t-s)}
\left(1+{\rm O}\left((s\vec{q}^2)^{-1},(t-s)/s\right)\right)$. In any case,
we recover the fluctuation-dissipation theorem
$T\wht{R}_{\rm eq}(t,s;\vec{q})
=\partial\wht{C}_{\rm eq}(t,s;\vec{q})/\partial s$
in the equilibrium regime as it should be. 

Motivated from studies in spin glasses, it is sometimes attempted to
separate correlation and response functions into an equilibrium and an
`ageing' part, viz. $C=C_{\rm eq} + C_{\rm age}$, $R=R_{\rm eq}+R_{\rm age}$. 
In our case, we would have
\BEA
\wht{R}_{\rm age}(t,s;\vec{q}) = \left(\frac{k(t)}{k(s)}-1\right)
\exp\left(-{\vec q}^{2}(t-s)\right)
\nonumber \\
\wht{C}^{0}_{\rm age}(t,s;\vec{q}) = 
\left( \wht{a}(\vec{q})-\frac{T}{{\vec{q}}^{2}}\right)
\exp\left(-{\vec{q}}^{2}(t+s)\right)
\label{eqage}
\EEA 
For $v(t)=0$, there is no `ageing' part in the response function.  

{}From these expressions, we can already extract a few general properties of 
the ageing process. First, for systems quenched to below 
their critical temperature $T_c>0$ and described by a MSR Gaussian 
action (\ref{eqMSRgen}), it is known from the dynamical renormalization group 
that the final temperature $T<T_c$ is an irrelevant parameter
and furthermore $T\to 0$ under renormalization \cite{Bray94,Bray00}. 
Then the long-time dynamics should be driven by the initial fluctuations, in
agreement with eq.~(\ref{eqage}) with $T=0$.  
On the other hand, for a critical quench $T=T_c$, the situation is different 
in that both initial and thermal fluctuation may contribute to the
long-time dynamics.  From eq.~(\ref{eqage}) we expect that the
small-$\vec{q}$ behaviour of the term $\wht{a}(\vec{q})-T_c/{\vec q}^{2}$ will 
determine the long-time dynamics.

%%%%%%%%%%%%%%%%%%%%%%%%%%%%%%%%%%%%%%%%%%%%%%%%%%%%%%%%%%%%%%%%%%%%%%%%%%%%%%%%
\section{Consequences of local scale-invariance of noiseless theories}
%%%%%%%%%%%%%%%%%%%%%%%%%%%%%%%%%%%%%%%%%%%%%%%%%%%%%%%%%%%%%%%%%%%%%%%%%%%%%%%%

\subsection{MSR formulation}

We now generalize the results of the previous section to any theory whose 
noiseless MSR action is Schr\"odinger-invariant. As we shall stay within
the context of classical field theory, the symmetries of the MSR action are
the same as for the corresponding Langevin equation (\ref{2:eq1}). 
The formulation of the
MSR action, using a non-conserved order parameter described by model A
dynamics \cite{Hohe77} is almost unchanged with respect to section~3. We have
\BEQ
S[\phi,{\wit \phi}] = S_{0}[\phi,{\wit \phi}]+{\cal S}[\phi, {\wit \phi}]
+\sigma[\phi, {\wit \phi}]
\label{eqa}
\EEQ        
where ${\cal S}[\phi, {\wit \phi}]$ and $\sigma[\phi, {\wit \phi}]$ 
are given by eqs.~(\ref{eqboun2},\ref{eqbulktherm}) and 
\BEQ
S_{0}[\phi,{\wit \phi}] = \int\!\D\vec{r}\,\D t\: 
{\wit \phi}\left(\frac{\partial \phi}{\partial t}
+\frac{\delta H}{\delta \phi}\right)
\EEQ
We shall assume throughout that {\em the noiseless action $S_0$ is 
Schr\"odinger-invariant} and that it includes an external time-dependent
potential $v=v(t)$. For a local effective potential $\cal H$, it is known
that spatial translation-invariance, dilatation-invariance (or dynamical
scaling) and Galilei-invariance are sufficient for having Schr\"odinger
invariance \cite{Henk03}. Therefore, from now on the dynamical exponent $z=2$.

In order to study the effects of the noise, we first discuss what becomes of 
the Galilei invariance of the noiseless theory. 
If the order parameter $\phi$ and the response field $\wit{\phi}$ are
quasiprimary, they should transform under a Galilei transformation
$t\mapsto t'=t$ and $\vec{r}\mapsto\vec{r}'=\vec{r}-\vec{v}t$ as 
(see eq.~(\ref{eqfg}))
\BEA
{\phi}'(t',{\vec r}') &=& f_{\vec{v}}(t,\vec{r})\phi(t,{\vec r})
\nonumber \\
{\wit \phi}'(t',{\vec r}') &=& 
f_{\vec{v}}^{-1}(t,\vec{r}){\wit \phi}(t,{\vec r})
\EEA 
where the companion function $f_{\vec{v}}$ reads \cite{Nied72}
\BEQ
f_{\vec{v}}(t,\vec{r}) = \exp\left[{\cal M}{\vec r}\cdot{\vec v}
-\frac{{\cal M}}{2}{\vec v}^{2}t\right]
\EEQ
The noisy contributions to the action transform as
\BEA
{\cal S}[\phi', {\wit \phi}']-{\cal S}[\phi, {\wit \phi}] 
&=& -T\int\!\D t\,\D\vec{r}\: {\wit \phi}^{2}(t,\vec{r})
\left(f_{\vec{v}}^{-2}(t,\vec{r}) -1 \right)
\nonumber \\
\sigma [\phi', {\wit \phi}']-\sigma [\phi, {\wit \phi}] 
&=& -\frac{1}{2}\int\!\D\vec{r}\,\D\vec{r}'a(\vec{r}-\vec{r}')
{\wit \phi}(0,\vec{r}){\wit \phi}(0,\vec{r}')
\left( f_{\vec{v}}^{-1}(0,\vec{r})f_{\vec{v}}^{-1}(0,\vec{r}')-1 \right)
\EEA
Therefore, for fixed temperature and initial conditions, 
the noise always destroys Galilei invariance.\footnote{The 
breaking of Galilei-invariance through thermal noise can be 
visualized as follows: 
consider a system in contact with a thermal bath at constant and
uniform temperature $T>0$. If the system moves with respect to the bath with
a constant speed $\vec{v}$, the apparent temperature measured in the system 
will depend on the angle between the direction of measurement and $\vec{v}$.}
Thus, the only dynamic symmetries of the noisy Langevin equation should be 
space translations and dilatations (and possibly space rotations). 

Consequently, one would merely expect the following scaling forms
\BEA
C(t,s,\vec{r},\vec{r}') &=& s^{-b}{\cal G}_{C}\left( \frac{t}{s}, 
\frac{(\vec{r}-\vec{r}')^{2}}{t-s} \right) 
\nonumber \\
R(t,s;\vec{r},\vec{r}') &=& s^{-a-1}{\cal G}_{R}\left( \frac{t}{s}, 
\frac{(\vec{r}-\vec{r}')^{2}}{t-s} \right) 
\label{eqscal}
\EEA
where ${\cal G}_{C,R}$ are undetermined scaling functions. For free-field 
theories, one would have $a=b=d/2-1$ by dimensional counting. 

\subsection{Response and correlation functions for the fluctuating theory}

The form of the scaling functions ${\cal G}_{R,C}$ will now be determined 
through a generalization of the expansions carried out in section~3. 
The main results will be eqs.~(\ref{eqR5}) and (\ref{eqC5}), (\ref{eqC6}),
respectively. 

\subsubsection{The two-point response function}

Consider a system which initially is at thermal equilibrium with temperature 
$T_i$ (which fixes the initial correlator $a(\vec{r})$) and quench it at
time $t=0$ to the final temperature $T=T_f$. The response function is still
given by eq.~(\ref{eqR1}) and the expansion of the 
exponential goes through as before.
Because of the Bargman superselection rule (\ref{3:gl:Barg}) which holds
because of the assumed Galilei invariance of $S_0$ only the lowest term
survives and we obtain
\BEQ 
R(t,s;\vec{r},\vec{r}')=R_{0}(t,s;\vec{r},\vec{r}')
\label{eqR4}
\EEQ
The expression of $R_{0}$ has been derived earlier and using the gauge 
transform (\ref{eqgauge}) we recover exactly the same result as in 
eq.~(\ref{eqres}), namely 
\BEQ
R\left( t,s;\vec{r},\vec{r}' \right) =
\delta_{x,\wit{x}}\: r_{0}\,\Theta(t-s) \left( t-s \right)^{-x}
\frac{k(t)}{k(s)}\exp \left( -\frac{\cal M}{2}
\frac{(\vec{r}-\vec{r}')^{2}}{(t-s)} \right) 
\label{eqR5}
\EEQ
We stress that, given only Galilei- and scale-invariance of the noiseless
theory, this result should hold for any initial and final temperature $T_i$
and $T_f$. 
At this stage, nothing has yet been said on the time-dependence of $v(t)$. 

\subsubsection{The two-point correlation function}

The two-point correlation function is found from eq.~(\ref{eqC2}). Again,
the arguments of section~3 go through and we find, using the explicit
expression (\ref{equni}) for the kernel $\kappa$ 
\BEQ
C(t,s;\vec{r},\vec{r}')=T \int\!\D u\,\D\vec{R}\: 
R_{0}^{(3)}(t,s,u;\vec{r},\vec{r}',\vec{R})+\frac{1}{2}
\int\!\D\vec{R}\,\D\vec{R}'\: a(\vec{R}-\vec{R}')
R_{0}^{(4)}(t,s,0,0;\vec{r},\vec{r'},\vec{R},\vec{R}') 
\label{eqC4}
\EEQ
where $R_{0}^{(3)}$ is the following three-point response function 
\BEQ 
R_{0}^{(3)}(t,s,u;\vec{r},\vec{r}',\vec{R}) := \left\langle\phi(t;\vec{r})
\phi(s;\vec{r}){\wit \phi}(u;\vec{R})^2\right\rangle_0
\label{eqR03}
\EEQ
and $R_0^{(4)}$ was already defined in eq.~(\ref{eqC2}). 
This central result will be the basis for all what follows. 
Consequently, the calculation of $C$ requires the computation of the 
{\em noiseless} three- and four-point functions $R_{0}^{(3)}$ and
$R_{0}^{(4)}$. This cannot entirely be done, since a general expression for
$R_{0}^{(4)}$ is not yet available. Of course, one might hope that 
through an extension of the Schr\"odinger algebra $\mathfrak{sch}_d$ to 
some infinite-dimensional Lie algebra techniques analogous to $2D$ conformal
invariance \cite{Bela84} might become applicable but the formulation of just
such an extension is an open problem.  

Here, we shall restrict to the case of vanishing initial correlations, that
is 
\BEQ \label{4:gl:aini}
a(\vec{R})=a_0\delta(\vec{R})
\EEQ
which corresponds to an infinite initial temperature $T_i=\infty$ and
where $a_0$ is a normalization constant.  
In fact, from renormalization group arguments the long-time behaviour of any 
system which is prepared in the high-temperature or paramagnetic phase should 
be described by this initial condition \cite{Bray94,Bray00}. 
Then
\BEA
C(t,s) &=& T\int \!\D u\,\D\vec{R}\: R_{0}^{(3)}(t,s,u;\vec{R})
+\frac{a_0}{2}\int \!\D\vec{R}\: R_{0}^{(3)}(t,s,0;\vec{R})
\label{eqC5}
\\
R_{0}^{(3)}(t,s,u;\vec{r}) &:=&
R_{0}^{(3)}(t,s,u;\vec{y},\vec{y},\vec{r}+\vec{y}) 
\: = \: \left\langle\phi(t;\vec{y})
\phi(s;\vec{y}){\wit \phi}(u;\vec{r}+\vec{y})^2\right\rangle_0
\EEA
Here, the field $\wit{\phi}^{\,2}$ is a composite field with mass $-2{\cal M}$ 
and scaling dimension $2\wit{x}_2$. Only for free fields, one has 
$\wit{x}_2 =\wit{x}$. 

Now, the three-point function of a Schr\"odinger-invariant theory with $v(t)=0$ 
is well-known since a long time \cite{Henk94,Henk02}. Denoting by $\cal R$ 
the response function in the case where $v(t)=0$, we have  
\BEA
{\cal R}_{0}^{(3)}(t,s,u;\vec{r}) &=& {\cal R}_{0}^{(3)}(t,s,u)
\exp\left[-\frac{{\cal M}}{2}\frac{t+s-2u}{(s-u)(t-u)}{\vec{r}}^{2} \right]
\Psi\left(\frac{t-s}{(t-u)(s-u)}{\vec{r}}^{2}\right)
\nonumber \\
{\cal R}_{0}^{(3)}(t,s,u) &=& \Theta(t-u)\Theta(s-u)\,
\left(t-u\right)^{-{\wit x}_2}
\left(s-u\right)^{-{\wit x}_2}\left(t-s\right)^{-x+{\wit x}_2}
\label{eqG03}
\EEA
where $\Psi$ is an arbitrary scaling function and $\cal M$ is the `mass' of the
field $\phi$. This result is brought to the case at hand through the 
gauge transformation (\ref{eqgauge}) and we find
\BEQ
R_{0}^{(3)}(t,s,u;\vec{r})=\frac{k(t)k(s)}{k^{2}(u)}
{\cal R}_{0}^{(3)}(t,s,u;\vec{r})
\label{eqR6}
\EEQ                    
Combining (\ref{eqC5}) and (\ref{eqR6}) we obtain, for the first time, 
a generic prediction for the form of the two-point correlation function. 
It is remarkable that under the condition (\ref{4:gl:aini}) only the noiseless 
three-point response functions are required in order to predict {\em any} 
two-time autocorrelator. 

It is useful to write down the autocorrelation function in the form
$C(t,s)=C_{th}(t,s)+C_{pr}(t,s)$. Here the thermal part 
$C_{th}$ and the preparation part $C_{pr}$ are given by
\BEA
C_{th}(t,s) &=& Ts^{d/2+1-x-{\wit x}_2}
\left(\frac{t}{s}-1\right)^{{\wit x}_2-x-d/2}
\int_{0}^{1}\!\,\D \theta\: \frac{k(t)k(s)}{k^{2}(s\theta)}
\left[\left(\frac{t}{s}-\theta\right)
\left(1-\theta\right)\right]^{d/2-{\wit x}_2}
\Phi\left(\frac{t/s+1-2\theta}{t/s-1}\right)
\nonumber \\
C_{pr}(t,s) &=& \frac{a_0}{2}
\frac{k(t)k(s)}{k^{2}(0)}s^{d/2-{\wit x}_2-x}
\left(\frac{t}{s}\right)^{d/2-{\wit x}}
\left(\frac{t}{s}-1\right)^{{\wit x}_2-x-d/2}
\Phi\left(\frac{t/s+1}{t/s-1}\right)
\nonumber \\
\Phi(w) &:=& \int\!\D\vec{R}\:\exp\left[-\frac{{\cal M}w}{2}\,{\vec R}^2\right]
\Psi\left({\vec R}^{2}\right)
\label{eqC6}
\EEA   
and we have explicitly used $s<t$.

As they stand, the above expressions for $C(t,s)$ do not yet necessarily
describe a dynamical scaling behaviour, since the form of $k(t)$ is still
completely general. We now assume in addition that we are dealing with a
system with dynamical scaling. In order to reproduce the usual 
phenomenology of ageing systems, we must have, at least for sufficiently
large times (see section 2)
\BEQ \label{4:gl:k}
k(t) \simeq k_0 t^{\digamma}
\EEQ
Comparing the general form of $R(t,s)$ as given in eq.~(\ref{eqR5}) with
the phenomenologically expected scaling
eqs.~(\ref{1:gl:SkalCR},\ref{1:gl:lambdaCR}), we read off
\BEQ \label{4:gl:xdg}
x = 1+a \;\; , \;\; \digamma = 1+a -\frac{\lambda_R}{2}
\EEQ
and in particular, the scaling function (\ref{1:gl:fR}) is recovered. 
On the other hand, for $C(t,s)$ we find the following scaling form, written
down separately for the thermal and the initial term, 
where $y=t/s\geq 1$ is the scaling variable
\BEA
C_{th}(t,s) &=& Ts^{-b_{th}}f_{C}^{\rm th}(y)
\nonumber \\
C_{pr}(t,s) &=& s^{-b_{pr}}f_{C}^{\rm pr}(y)
\label{eqC7}
\EEA
where the scaling functions $f_{C}^{\rm pr}$ and $f_{C}^{\rm th}$ are given by
\BEA
f_{C}^{\rm th}(y) &=& y^{\digamma}\left(y-1\right)^{{\wit x}_2-x-d/2}
\int_{0}^{1}\!\,\D \theta\: 
\theta^{-2\digamma}\left[\left(y-\theta\right)
\left(1-\theta\right)\right]^{d/2-{\wit x}_2}
\Phi\left(\frac{y+1-2\theta}{y-1}\right)
\nonumber \\
f_{C}^{\rm pr}(y) &=& \frac{a_0}{2} y^{d/2-{\wit x}_2+\digamma}
\left(y-1\right)^{{\wit x}_2-x-d/2}
\Phi\left(\frac{y+1}{y-1}\right)
\label{eqscfunc}
\EEA
with the non-equilibrium exponents 
\BEA
b_{th} &=& x+{\wit x}_2-1-d/2
\nonumber \\
b_{pr} &=& x+{\wit x}_2-2\digamma-d/2
\label{eqbexp} 
\EEA
Provided that $\Phi(1)$ is finite, the asymptotic behaviour of the scaling
functions for $y$ large can be worked out. We expect 
$f_{C}(y)\sim y^{-\lambda_{C}/2}$ and find
\BEQ
\lambda_C^{th} = \lambda_C^{pr} = 2(x-\digamma)
\EEQ
Therefore, comparing this with (\ref{4:gl:xdg}), we have shown: \\ ~\\
{\it For any system with an infinite-temperature initial state 
(\ref{4:gl:aini}), quenched to a temperature $T< T_c$ and whose 
noiseless part is locally scale-invariant with $z=2$, one has}
\BEQ \label{4:gl:ll}
\lambda_C = \lambda_R . 
\EEQ

\noindent 
For non-equilibrium critical dynamics (that is $T=T_c$) the same conclusion
can be drawn if after renormalization one still has $z=2$. 
While eq.~(\ref{4:gl:ll}) certainly agrees with the evidence available 
from models studied either analytically or numerically, we are not aware of any 
other general proof of this equality between the autocorrelation and 
autoresponse exponents for a fully disordered initial state. 
 
In sections~5-7, we shall present extensive tests of the prediction 
(\ref{eqC7},\ref{eqscfunc},\ref{eqbexp}) for $C$ and 
(\ref{eqR5},\ref{4:gl:xdg}) for $R$, respectively. The main hypothesis going 
into it is the requirement of Galilei-invariance of the noiseless theory, 
while the other conditions appear to be habitually admitted in the description 
of physical ageing. 

\subsubsection{Autocorrelation function in phase-ordering kinetics}

In order to understand the result (\ref{eqC7}) for $C(t,s)$ better, 
we now study the
two contributions separately. First, we consider the `preparation' part
$C_{pr}$. This term is expected to describe the late-time behaviour
of a system quenched to a temperature $T<T_c$. Indeed, renormalization-group
arguments show \cite{Bray94} that in this case $T$ is an irrelevant variable
and is renormalized towards zero. Then the thermal contribution $C_{th}$
vanishes. Therefore, the non-equilibrium exponent $b=b_{pr}$ is read off from
eq.~(\ref{eqbexp}). In addition, we know that $b=0$ in the low-temperature
phase. This implies
\BEQ
\wit{x}_2 - x = \frac{d}{2} - \lambda_C \leq 0 
\EEQ
because of a well-known inequality \cite{Yeun96}. Only for free fields, one
has $\lambda_C=d/2$, otherwise $\wit{x}_2$ is a new nontrivial exponent. 
In the scaling limit, we thus have $C(t,s)=f_{C}(t/s)$ where 
\BEQ \label{eqC11}
f_{C}(y) = \frac{a_0}{2} y^{\lambda_{C}/2}\left(y-1\right)^{-\lambda_C}
\Phi\left(\frac{y+1}{y-1}\right)
\EEQ 

The form of $f_C(y)$ still depends on the unknown function $\Phi(w)$ which
in turn depends on $\Psi(\rho)$, see (\ref{eqC6}). 
We attempt to fix its form and reconsider the noiseless response function
$R_0^{(3)}(t,s,0;\vec{r})$ which describes a response of the 
autocorrelation $C(t,s)=\langle\phi(t)\phi(s)\rangle$. It appears 
to be a reasonable requirement that there should be no singularity in
$R_0^{(3)}$ when $t=s$. Using the explicit form eqs.~(\ref{eqG03},\ref{eqR6})
this leads to the following limit behaviour
\BEQ
\Psi(\rho) \simeq \Psi_0 \rho^{\lambda_C-d/2} \;\; ; \;\; \rho\to 0
\EEQ
where $\Psi_0$ is a constant. 
If this leading term should be still accurate for larger values of $\rho$,
the following expression for the scaling function $\Phi(w)$ is found
\BEQ \label{eqPhi}
\Phi(w) \approx \Psi_0 S_d \frac{\Gamma(\lambda_C)}{2}
\left(\frac{2}{\cal M}\right)^{\lambda_C} \cdot w^{-\lambda_C} 
=: \Phi_0 \cdot w^{-\lambda_C} 
\EEQ
where $S_d$ is the surface of the unit sphere in $d$ dimensions. 
Eq.~(\ref{eqPhi}) should hold in the $w\to\infty$ limit. 
Provided this from is still valid for all $w$, we would obtain the 
following simplified form, with $y=t/s$ 
\BEQ \label{eqC12}
C(t,s)\approx\frac{a_0\Phi_0}{2}
\left(\frac{(y+1)^2}{y}\right)^{-\lambda_C/2}
= M_{\rm eq}^{2}\, 
\left(\frac{(y+1)^2}{4y}\right)^{-\lambda_C/2}
\;\; ; \;\; T=0
\EEQ
where we also related the leading constant to the squared equilibrium 
magnetization $M_{\rm eq}^2$, in order to recover the usual scaling form
$C(t,s)=M_{\rm eq}^2 f_C(t/s)$ with $f_C(1)=1$, see \cite{Godr02}.

\subsubsection{Autocorrelation function for critical dynamics}

Second, let us turn to the thermal part. It should dominate the autocorrelation
for quenches to the critical point $T=T_c$, given the initial condition
(\ref{4:gl:aini}) and under the assumption that $z=2$ even at criticality. 
{}From eq.~(\ref{eqbexp}), the exponent $b=b_{th}$ can be read off and
we have 
\BEQ
\wit{x}_2 = b -a +\frac{d}{2}
\EEQ
Under the stated conditions, the preparation term drops out at large times
and we find 
\BEA
C(t,s) &=& T_{c}s^{-b}f_{C}(t/s)
\nonumber \\
f_{C}(y)  &=& y^{\digamma}\left(y-1\right)^{b-2a-1}
\int_{0}^{1}\!\,\D \theta\: \theta^{-2\digamma}
\left[(y-\theta)(1-\theta)\right]^{a-b}
\Phi\left(\frac{y+1-2\theta}{y-1}\right)
\label{eqC9}
\EEA     

At criticality, one expects the following relationship between the 
nonequilibrium exponents $a$ and $b$ 
\BEQ
a=b=\frac{2\beta}{\nu z}=\frac{d-2+\eta}{z}
\label{eqexp2}
\EEQ
where $\beta,\nu,\eta$ are well-known equilibrium critical exponents
(see section~1). 
To understand eq.~(\ref{eqexp2}), recall that $C(s,s)\sim s^{-b}$. 
On the other hand, from the space-time scaling $|\vec{r}|^z\sim t$, 
one expects the equilibrium correlator to decay as 
$C_{\rm eq}\sim |\vec{r}|^{-bz}$ and the second equality in 
eq.~(\ref{eqexp2}) follows. Finally, $a=b$ is a necessary condition for having
a non-vanishing limit fluctuation-dissipation ratio $X_{\infty}$. 

Now, if we let $a=b$ in (\ref{eqC9}), we find
\BEQ \label{eqC15}
f_C(y) = y^{\digamma} \left( y-1\right)^{-1-a} 
\int_{0}^{1}\!\,\D \theta\: \theta^{-2\digamma}
\Phi\left(\frac{y+1-2\theta}{y-1}\right)
\EEQ
which is the most general form compatible with the standard phenomenological
constraints. An approximate form of the scaling function $\Phi(w)$ may be 
obtained from the requirement that the three-point response function
${\cal R}_0^{(3)}(t,s,u;\vec{r})$ is non-singular for $t=s$. We now have
$\wit{x}_2=d/2$ and find $\Psi(\rho)\simeq \Psi_{0,c}\,\rho^{x-d/2}$ as 
$\rho\to 0$. This leads to, for $w\to\infty$
\BEQ \label{Phi0c}
\Phi(w) \approx \Phi_{0,c}\, w^{-1-a}
\EEQ
where $\Phi_{0,c}$ is a constant. 
If in addition we may use this form also for finite values of $w$,
we would obtain the simplified form
\BEQ \label{eqC16}
f_C(y) \approx \Phi_{0,c}\, y^{1+a-\lambda_C/2}
\int_{0}^{1}\!\,\D \theta\: \theta^{\lambda_C-2-2a}
\left( y+1-2\theta\right)^{-1-a}
\;\; ; \;\; T=T_c
\EEQ

Summarizing, the phenomenological comparison of the autocorrelation function, 
as predicted by Schr\"odinger-invariance, and assuming a totally disordered 
initial state, with simulational or experimental data
will be based on eqs.~(\ref{eqC11}) and (\ref{eqC15}) for quenches to 
$T<T_c$ and $T=T_c$, respectively. In full generality this will allow to 
obtain information on the scaling function $\Phi(w)$. If in addition 
the heuristic idea of the absence of singularities at $t=s$ in the
three-point response function $R_0^{(3)}$ should be sufficient to fix the
form of this response function, the simplified forms 
(\ref{eqC12},\ref{eqC16}) apply and the scaling function $f_C(y)$
is completely specified in terms of the exponents $a$ and $\lambda_C$.

%%%%%%%%%%%%%%%%%%%%%%%%%%%%%%%%%%%%%%%%%%%%%%%%%%%%%%%%%%%%%%%%%%%%%%%%%%%%%%%%
\section{Tests of local scale-invariance in exactly solvable models}
%%%%%%%%%%%%%%%%%%%%%%%%%%%%%%%%%%%%%%%%%%%%%%%%%%%%%%%%%%%%%%%%%%%%%%%%%%%%%%%%

In this and the next two sections we describe phenomenological tests of the 
predictions of Schr\"odinger-invariance, that is local scale-invariance with
$z=2$, which were derived in section 4 in concrete physical models of ageing 
behaviour. 

\subsection{Kinetic spherical model}

The kinetic spherical model is often formulated in a field-theoretic
fashion as the $n\to\infty$ limit of the O($n$)-symmetric vector model. 
For our purposes, it is more useful to start directly from a lattice system
and to take the continuum limit 
later, following \cite{Cugl95,Godr00b,Zipp00,Pico02,Pico03,Paes03}. 

Consider a hypercubic lattice with $\cal N$ sites. 
At each site $\vec{r}$ there 
is a real time-dependent variable $\phi(t,\vec{r})$ 
such that the mean spherical constraint
\BEQ \label{6:moyen}
\left\langle \sum_{\vec{r}} \phi(t,\vec{r})^2 \right\rangle = {\cal N}
\EEQ
holds. The Hamiltonian is 
${\cal H} = - \sum_{<\vec{r},\vec{r}'>} \phi(t,\vec{r})\phi(t,\vec{r}')$
where the sum extends over pairs of nearest neighbour sites. The
(non-conserved) dynamics is given in terms of a Langevin equation of the type
eq.~(\ref{2:eq1})
\BEA
\frac{\partial\phi(t,\vec{r})}{\partial t} 
&=& \sum_{\vec{n}(\vec{r})} \phi(t,\vec{n}) 
- (2d+v(t))\phi(t,\vec{r}) +\eta(t,\vec{r}) \label{6:kinSM}
\\
&\simeq& \Delta \phi(t,\vec{r}) - v(t) \phi(t,\vec{r}) +\eta(t,\vec{r})
\EEA
where $\vec{n}(\vec{r})$ runs over the nearest neighbours 
of the site $\vec{r}$.
In the second line, a formal continuum limit was taken (for simplicity, all
rescalings with powers of the lattice constant $a$ were suppressed). Finally,
$\eta(t,\vec{r})$ is the usual Gaussian noise (see section~1) and the Lagrange
multiplier $v(t)$ is fixed such that the mean spherical constraint is
satisfied.\footnote{A careful study shows that provided the limit 
${\cal N}\to\infty$ is taken {\em before} any long-time limit, the mean
spherical constraint (\ref{6:moyen}) and a full, non-averaged, spherical  
constraint lead to the same results \cite{Fusc03}.} Therefore, the kinetic
spherical model perfectly fits into the context of a Schr\"odinger equation
in a time-dependent potential $v=v(t)$ discussed in section 3.
We therefore expect that the free-field predictions 
eqs.~(\ref{eqR2},\ref{eqcornoise}) will be fully confirmed. 

In order to see this explicitly, we recall the elements of the exact solution
of the Langevin equation (\ref{6:kinSM}). If we set
\BEQ 
g(t) = \exp\left( 2 \int_{0}^{t} \!\D u\, v(u)\right)
\EEQ
it can be shown \cite{Cugl95,Godr00b,Pico02} that $g(t)$ 
is the unique solution of the Volterra integral equation
\BEQ \label{6:Volt}
g(t) = A(t) + 2T \int_{0}^{t} \!\D t'\, f(t-t') g(t')
\EEQ
where $g(0)=1$ and 
\BEA
f(t) &=& \Theta(t) \left( e^{-4t} I_0(4t)\right)^d 
\nonumber \\
A(t) &=& \frac{1}{(2\pi)^d} \int_{\cal B} \!\D\vec{q}\: 
\sum_{\vec{r}} a(\vec{r}) e^{-2\omega(\vec{q})t-\II\vec{q}\cdot\vec{r}}
\EEA
and $\omega(\vec{q})=2\sum_{i=1}^{d} (1-\cos q_i)$ is the lattice dispersion
relation, $\cal B$ the Brillouin zone while $I_0$ is a 
modified Bessel function.
We stress that because of eq.~(\ref{6:Volt}), $g(t)$ does {\em not} depend
on the order parameter $\phi(t,\vec{r})$.  

We now show that the prediction (\ref{eqR2},\ref{eqcornoise}) for the
two-point functions can be fully reproduced. 
We begin with the response function. In the spherical model, it is exactly 
given by \cite{Godr00b,Pico02}
\BEA
R(t,s;\vec{r},\vec{r}') &=& 
\left.\frac{\delta\langle\phi(t,\vec{r})\rangle}{\delta h(s,\vec{r}')}
\right|_{h=0} 
\nonumber \\
&=& \left[ \prod_{i=1}^{d} e^{-2(t-s)} 
I_{\vec{r}_i-\vec{r}_{i}'}\left( 2(t-s)\right)\right]
\sqrt{\frac{g(s)}{g(t)}\,} 
\nonumber \\
&\simeq& \left( 4\pi(t-s)\right)^{-d/2} 
\exp\left(-\frac{(\vec{r}-\vec{r}')^2}{4(t-s)}\right)\,
\exp\left(-\int_{s}^{t}\!\D u\, v(u)\right)\, \Theta(t-s)
\label{6:gl:Rexplizit}
\EEA
where in the second line the limit $t-s\gg 1$ was taken and $I_r$ is a 
modified Bessel function. In particular, this reproduces the known
autoresponse function \cite{Godr00b}
\BEQ 
R(t,s)=R(t,s;\vec{r},\vec{r})=f((t-s)/2)\sqrt{g(s)/g(t)}
\EEQ
This is in exact agreement with eq.~(\ref{eqR2}) and
we identify the mass ${\cal M}=1/2$. 

We now turn to the correlator but the sake of brevity merely deal with the
autocorrelator explicitly. 
In the spherical model one has (with $t>s$) \cite{Godr00b,Pico02,Paes03}
\BEQ \label{eq:SM:C}
C(t,s) = \left\langle\phi(t,\vec{r})\phi(s,\vec{r})\right\rangle = 
\frac{1}{\sqrt{g(t)g(s)\,}}
\left[ A\left(\frac{t+s}{2}\right) + 2T \int_{0}^{s}\!\D u\, 
f\left(\frac{t+s}{2}-u\right) g(u) \right]
\EEQ
In order to show how to recover this explicitly from eq.~(\ref{eqcornoise}), we
write again $C(t,s)=: C_{pr} + C_{th}$ and discuss the two terms separately. 
The first one is, where we use the explicit form (\ref{6:gl:Rexplizit}) 
of $R=R_0$ 
\BEA
C_{pr} &=& \int_{\mathbb{R}^{2d}} \!\D\vec{y}\D\vec{y}'\: 
\left( 4\pi t \, 4\pi s\right)^{-d/2} 
\sqrt{ \frac{g(0)}{g(t)}\frac{g(0)}{g(s)}\,} 
\exp\left[ -\frac{(\vec{r}-\vec{y})^2}{4t} - \frac{(\vec{r}-\vec{y}')^2}{4s}
\right] 
\nonumber \\
&=& \frac{(2\pi)^{-d/2}}{\sqrt{g(t)g(s)}} 
\int_{\mathbb{R}^{2d}} \!\D\vec{q}\D\vec{q}'\: 
e^{-\vec{q}^2t-\vec{q}'^2s+\II (\vec{q}+\vec{q}')\cdot\vec{r}} \cdot J_d
\EEA
where
\BEQ
J_d := \int_{\mathbb{R}^{2d}} \!\D\vec{y}\D\vec{y}'\:
a(\vec{y}-\vec{y}')\,e^{-\II\vec{q}\cdot\vec{y}-\II\vec{q}'\cdot\vec{y}'}
\EEQ
and we used $g(0)=1$. In order to calculate $J_d$, we set
$\vec{\zeta} = \vec{y}- \vec{y}'$, $\vec{\zeta}' = \vec{y}+ \vec{y}'$.
Then the Jacobian 
$\left|\partial(\vec{y},\vec{y}')/\partial(\vec{\zeta},\vec{\zeta}')\right|
=2^{-d}$ and we have
\BEA
J_d &=& 2^{-d} \int_{\mathbb{R}^{2d}} \!\D\vec{\zeta}\D\vec{\zeta}'\: 
a(\vec{\zeta})\,  
e^{-\II\frac{\vec{\zeta}'}{2}\cdot(\vec{q}+\vec{q}')}
e^{-\II\frac{\vec{\zeta}}{2}\cdot(\vec{q}-\vec{q}')}
\nonumber \\
&=& (2\pi)^d \delta(\vec{q}+\vec{q}') 
\int_{\mathbb{R}^d} \!\D\vec{\zeta}\: a(\vec{\zeta}) 
e^{-\II\vec{q}\cdot\vec{\zeta}}
\EEA
We finally obtain
\BEQ
C_{pr} = \frac{1}{\sqrt{g(t)g(s)}}\frac{1}{(2\pi)^d} 
\int_{\mathbb{R}^d} \!\D\vec{q}\: e^{-\vec{q}^2(t+s)}\,
\int_{\mathbb{R}^d} \!\D\vec{\zeta}\: a(\vec{\zeta})\, 
e^{-\II\vec{q}\cdot\vec{\zeta}}
\EEQ
When the support of $a(\vec{\zeta})$ is restricted to the hypercubic lattice,
we therefore have indeed for long times $t+s\gg 1$
\BEQ
C_{pr} \simeq \frac{A\left((t+s)/2\right)}{\sqrt{g(t)g(s)}}
\EEQ
in agreement with the first term in (\ref{eq:SM:C}). 
The second term in (\ref{eqcornoise}) is analyzed as follows, for $t>s$
\BEA
C_{th} &=& 2T \int_{0}^{\infty}\!\D u \int_{\mathbb{R}^d} \!\D\vec{y}\: 
\left( 4\pi (t-u)\, 4\pi (s-u)\right)^{-d/2} 
\sqrt{\frac{g(u)}{g(t)}\frac{g(u)}{g(s)}}\,
\nonumber \\
& & \times 
\exp\left[ - \frac{(\vec{r}-\vec{y})^2}{4(t-u)} 
- \frac{(\vec{r}-\vec{y})^2}{4(s-u)}\right] \,
\Theta(t-u) \Theta(s-u) 
\nonumber \\
&=& 2T\int_{0}^{s}\!\D u \int_{\mathbb{R}^{3d}}\!\D\vec{q}\D\vec{q}'\D\vec{y}\:
e^{-\vec{q}^2t-\vec{q}'^2s+(\vec{q}^2+\vec{q}'^2)u 
+\II(\vec{q}+\vec{q}')\cdot\vec{r}} \,
e^{-\II(\vec{q}+\vec{q}')\cdot\vec{y}}\,
\frac{g(u)}{\sqrt{g(t)g(s)}}
\nonumber \\
&=& \frac{2T}{\sqrt{g(t)g(s)}} \int_0^s\!\D u\, g(u)\: (2\pi)^{-d}
\int_{\mathbb{R}^d}\!\D\vec{q}\: e^{-2\vec{q}^2\left( (t+s)/2-u\right)}
\nonumber \\
&\simeq& \frac{2T}{\sqrt{g(t)g(s)}} \int_0^s\!\D u\, g(u) 
f\left(\frac{t+s}{2}-u\right)
\EEA
where the last line holds for sufficiently large arguments of the function
$f$. Taken together with $C_{pr}$, the expected agreement between the
general result eq.~(\ref{eqcornoise}) and the exact expression 
(\ref{eq:SM:C}) for the spherical model is thus recovered. The full space-time 
correlator can be checked similarly. 

It is interesting to note that for the confirmation of $R$, see
(\ref{6:gl:Rexplizit}), we need the condition $t-s\gg 1$, while for the 
confirmation of $C$, we also need $s\gg 1$ (which implies $t+s\gg 1$). 
%%MH the further condition $t+s\gg 1$ is required as well. 

So far, we have not yet used the explicit form of $g(t)$ which follows from
eq.~(\ref{6:Volt}). Indeed, it is well-known that for long times, one has
\cite{Godr00b,Pico02,Henk02}
\BEQ
g(t) \sim t^{-2\digamma}
\EEQ
For the fully disordered initial conditions (\ref{4:gl:aini}), the 
exponent $\digamma$ takes the following values: (i) for $T<T_c$, one has 
$\digamma=d/4$ and (ii) for $T=T_c$, one has $\digamma=1-d/4$ if $2<d<4$ and
$\digamma=0$ if $d>4$. This is exactly the form expected from a
potential of the form $v(t)\simeq \digamma/t$, see eq.~(\ref{eqsc}). 

It is instructive to compare also the explicit result for the two-point 
functions with the expectations coming from local scale-invariance. 
For the two-time response function, 
this confirmation has already been carried out for both
the autoresponse function $R(t,s)$ as well as the space-time response
$R(t,s;\vec{r})$ \cite{Henk01,Henk02,Pico02} and need not be repeated here. 
Therefore we concentrate on the two-time autocorrelation function $C(t,s)$. 
First, we consider the case $T<T_c$ where from the exact solution it is 
well-known that \cite{Godr00b,Newm90}
\BEQ
C(t,s) = M_{\rm eq}^2 f_C(t/s) \;\; , \;\; 
f_C(y) = \left( \frac{(y+1)^2}{4y} \right)^{-d/4}
\EEQ
and we read off from the $y\to\infty$ limit the exponent $\lambda_C=d/2$. 
Clearly, this exact result is in full agreement with the prediction 
(\ref{eqC12}) of local scale invariance. Second, we consider the case $T=T_c$. 
Then the exact solution gives \cite{Godr00b}
\BEQ \label{5:gl:Csm}
\renewcommand{\arraystretch}{2.0}
C(t,s) = (4\pi)^{-d/2} T_c s^{-d/2+1} f_C(t/s) \;\; , \;\;
f_C(y) = \left\{ {{ \begin{array}{*{20}ll}
\frac{4}{d-2} (y-1)^{-d/2+1} y^{1-d/4} (y+1)^{-1} & \mbox{\rm ~~;~if $2<d<4$}\\
\frac{2}{d-2}\left[ (y-1)^{-d/2+1} - (y+1)^{-d/2+1}\right] & 
\mbox{\rm ~~;~if $d>4$}
\end{array} }}\right.
\renewcommand{\arraystretch}{1.0}
\EEQ
{}From this, we read off the exponents $a=b=d/2-1$ for all $d>2$ and furthermore
$\lambda_C=3d/2-2$ for $2<d<4$ and $\lambda_C=d$ for $d>4$, respectively. 
Inserting the exponent values into (\ref{eqC16}) it is straightforward to
show that eq.~(\ref{5:gl:Csm}) is indeed reproduced. 

In particular, we see that in the spherical model the asymptotic forms
eqs.~(\ref{eqPhi},\ref{Phi0c}), respectively, are indeed exact as should be
expected for a free-field theory. 

In conclusion, the exact results of the kinetic spherical
model for both two-time correlation and response functions are in full 
agreement with the predictions of Schr\"odinger invariance.

\subsection{XY model in spin-wave approximation}

Another system which can be exactly analysed is the kinetic XY model in
spin-wave approximation. As we shall see, it provides an instructive example
on the correct identification of the quasi-primary scaling fields in a given 
model. 

\subsubsection{Formulation and observables}

The XY model describes the interaction between planar spin variables 
\BEQ
\vec{S}(\vec{r}) = \cos(\phi(\vec{r})) \vec{e}_{1}
+\sin(\phi(\vec{r}))\vec{e}_{2} = 
\left( \begin{array}{l} \cos \phi(\vec{r}) \\ \sin \phi(\vec{r})
\end{array} \right)
\EEQ
which are attached to the sites $\vec{r}$ of a $d$-dimensional hypercubic 
lattice and $\phi(\vec{r})$ is the phase. The Hamiltonian is 
\BEQ
{\cal H}[\phi]=-\sum_{\langle \vec{r},\vec{r'} \rangle} 
\vec{S}(\vec{r})\cdot\vec{S}(\vec{r'})
=-\sum_{\langle \vec{r},\vec{r'} \rangle} 
\cos\left(\phi(\vec{r})-\phi(\vec{r'})\right)
\EEQ
where the coupling constant $J$ has been set to unity and 
the sum runs over nearest neighbours. The relaxational dynamics is assumed to
be described by a Langevin equation. We prepare the system initially
a temperature $T_i$ and quench it at time $t=0$ to the final 
temperature $T=T_{f}$, so that the angular variable obeys \cite{Bray94} 
\begin{equation}
\frac{\partial \phi(t,\boldsymbol{r})}{\partial t}  
= - \frac{\delta {\cal H}[\phi]}{\delta \phi(t,\boldsymbol{r})} +
\eta(t,\boldsymbol{r}).
\label{eqdyn}
\end{equation}
where $\eta$ represents an uncorrelated Gaussian noise with zero mean and 
variance
\BEQ
\langle \eta (t,\boldsymbol{r}) \eta(t',\boldsymbol{r'})\rangle
= 2T_{f} \delta(t-t')\delta(\vec{r}-\vec{r}')
\EEQ

Here, we shall exclusively study the coarsening dynamics in the low-temperature
regime, that is 
\BEQ
T_{i},T_{f} \ll T_{c}(d)
\EEQ 
where $T_{c}(d)$ is the critical temperature of the $XY$ model in $d$ 
dimensions (if $d=2$, $T_{c}(2)=T_{\rm KT}$ is the Kosterlitz-Thouless 
temperature of the transition). Then the so-called spin-wave approximation 
\cite{Rute95a,Bert01} can be used which amounts to expand 
$\cal H$ in powers of $\phi(\vec{r})-\phi(\vec{r'})$. Shifting the energy by 
a constant, the Hamiltonian reads to lowest order 
\BEQ \label{5:gl:HXYsw}
{\cal H}[\phi]= \frac{1}{2}\int \!\D\vec{r}\: 
\left(\vec{\nabla} \phi(\vec{r})\right)^{2}  
\EEQ   
In writing this, we have implicitly absorbed the spin-wave stiffness
\cite{Kost73,Rute95a} into a redefinition of the temperatures. 
In $2D$, it is known
that below $T_{\rm KT}$, any vortices present will be tightly bound and for
distances larger than the characteristic pair size, the XY model renormalizes
to the Hamiltonian (\ref{5:gl:HXYsw}) \cite{Rute95a}.\footnote{For quenches
from {\em above} $T_{\rm KT}$ in $2D$, vortex configurations also become 
important and this leads to logarithmic scaling, see \cite{Roja99} for details.}

We are interested in the properties of the two-point functions. It appears
natural to define two-time correlation and linear response functions in terms
of the magnetic variables 
\BEA
\Gamma(t,s;\vec{r},\vec{r}') &:=&  
\langle \vec{S}(t,\vec{r})\cdot\vec{S}(s,\vec{r}') \rangle 
=\langle \cos \left( \phi(t,\vec{r})-\phi(s,\vec{r}') \right) \rangle
\nonumber \\
\rho(t,s;\vec{r},\vec{r}') &:=& 
\lim_{\vec{h}\to \vec{0}}
\frac{\partial \langle \vec{S}(t,\vec{r}) \rangle}{\partial\vec{h}(s,\vec{r}')}
\label{eq:defS}
\EEA
where the response is found by adding a term 
$\delta{\cal H}_{\rm mag}=\sum_{\vec{r}} \vec{h}\cdot\vec{S}$ to 
$\cal H$. Alternatively, one may also consider the analogous quantities defined 
for the angular variables 
\BEA
C(t,s;\vec{r},\vec{r}') &:=& \langle \phi(t,\vec{r}) \phi(s,\vec{r}') \rangle
\nonumber \\
R(t,s;\vec{r},\vec{r}') &:=& \lim_{{h}\to \vec{0}}
\frac{\partial \langle \phi(t,\vec{r}) \rangle}{\partial{h}(s,\vec{r}')}
\label{eq:defP}
\EEA
(where a perturbation $\delta{\cal H}_{\rm ang}=\sum_{\vec{r}} h \phi$ 
should have been added). 

In order that the spin-wave approximation be applicable, we must start from
an (almost) ordered initial state of the system. Therefore, we require
the following initial value for the magnetic correlator, which reads in
Fourier space
\BEQ
\wht{a}(\vec{q}) = \wht{C}(0,0;\vec{q})
= \frac{2\pi\eta(T_{i})}{q^{2}}=\frac{T_{i}}{q^{2}}
\label{eqiniXY}
\EEQ 
where $\eta(T_{i})$ is the standard equilibrium critical exponent and the
the relation $2\pi\eta(T_{i})=T_{i}$ valid in the spin-wave approximation 
was used, see \cite{Kost73,Rute95a,Bert01}.

\subsubsection{Non-equilibrium statistical field theory}

As before, we introduce a Martin-Siggia-Rose formalism  
which characterizes the system is term of an action 
$S[\phi,{\wit \phi}, \vec{h}]$ depending on the phase field $\phi$ and an
associated response field $\wit{\phi}$ and we also include a
(possibly space-dependent) magnetic field $\vec{h}=\sum_{i}h_{i}\vec{e}_{i}$.
We decompose the action 
\BEQ
S[\phi, {\wit \phi}] = \Sigma[\phi, {\wit \phi}, \vec{h}]
+\sigma[\phi, {\wit \phi}]
\label{eqact}
\EEQ 
into a bulk term $\Sigma[\phi, {\wit \phi}, \vec{h}]$ and an initial term 
$\sigma[\phi, {\wit \phi}]$. These two terms include the thermal and the
initial noise. Explicitly
\BEQ
\Sigma[\phi, {\wit \phi}, \vec{h}]=
\int \!\D t\,\D\vec{r}\: \wit{\phi}\left[\frac{\partial \phi}{\partial t}
-\Delta \phi+\sin(\phi)h_{1}-\cos(\phi)h_{2}\right]
-T\int \!\D t\,\D\vec{r}\: {\wit \phi}(t,\vec{r}){\wit \phi}(t,\vec{r})
\label{eqbulkXY}
\EEQ  
and 
\BEQ
\sigma[\phi, {\wit \phi}]=-\frac{1}{2}\int \!\D\vec{r}\,\D\vec{r}'\:  
{\wit \phi}(0,\vec{r})a(\vec{r}-\vec{r}'){\wit{\phi}(0,\vec{r}')}
\label{eqbounXY}
\EEQ   
where the function $a(\vec{r})$ describes the initial conditions 
according to eq.~(\ref{eqiniXY}). 

We now simplify the general expressions for the two-point functions. There is
nothing to do for the angular correlation function 
$C(t,s;\vec{r},\vec{r}')=\langle \phi(t,\vec{r})\phi(s,\vec{r}')\rangle$ 
and we start with the magnetic correlation function $\Gamma$ which is given by
\BEQ
\Gamma(t,s;\vec{r},\vec{r}')=\int  \!{\cal D}\phi{\cal D}{\wit \phi}\:  
\cos \left(\phi(t,\vec{r})-\phi(s,\vec{r}')\right)  
\exp\left(-S[\phi, {\wit \phi}]\right) 
\label{eqcXY}
\EEQ
For a vanishing magnetic field the bulk action 
$\Sigma[\phi,{\wit \phi},\vec{0}]$ is
a quadratic form in the fields $\phi, {\wit \phi}$ which are therefore
Gaussian. Standard techniques explained in appendix~A lead to
\BEA
\Gamma(t,s;\vec{r},\vec{r}') &=&\exp \left[-\frac{1}{2}
\left\langle(\phi(t,\vec{r})-\phi(s,\vec{r}'))^{2}\right\rangle\right]
\\
&=& \exp \left[ C(t,s;\vec{r},\vec{r}')-
\frac{ C(t,t;\vec{r},\vec{r})+C(s,s;\vec{r}',\vec{r}')}{2} \right]
\label{eqXYcor}
\EEA
where in the second line the argument of the exponential was expanded.
This gives $\Gamma$ in terms of angular correlators $C$. 

Next, we consider the response functions. For the angular response $R$,
we quote from the MSR formalism the standard result
\BEQ
R(t,s;\vec{r},\vec{r}')= \langle \phi(t,\vec{r}){\wit \phi}(s,\vec{r}') 
\rangle 
\label{eqangres}
\EEQ 
It remains to consider the response of the spin vector $\vec{S}$ at time $t$ 
and position $\vec{r}$ to some magnetic field $\vec{h}(s,\vec{r}')$ 
at time $s$ and position $\vec{r}'$. From the definition (\ref{eq:defS}) we have
\BEQ
\rho(t,s;\vec{r},\vec{r}') = 
\lim_{\vec{h}\rightarrow 0}\left[ \frac{ \langle\langle 
\cos \phi(t,\vec{r})\rangle \rangle-\langle 
\cos \phi(t,\vec{r})\rangle}{h_{1}(s,\vec{r}')}
+\frac{\langle\langle \sin \phi(t,\vec{r})\rangle\rangle
-\langle\sin \phi(t,\vec{r})\rangle}{h_{2}(s,\vec{r}')} \right]
\EEQ
where the average $\langle\langle\cdot\rangle\rangle$ is to be taken with
a magnetic field. Expanding the action (\ref{eqact}) to first order in both 
components $h_{1},h_{2}$ of the magnetic field, we find
\BEA
\langle\langle \cos \phi(t,\vec{r})\rangle \rangle
=\langle \cos \phi(t,\vec{r})\rangle+h_{1}(s,\vec{r}')
\langle\cos\phi(t,\vec{r})\sin\phi(s,\vec{r}'){\wit\phi(s,\vec{r}')}\rangle
\nonumber \\
\langle\langle \sin \phi(t,\vec{r})\rangle \rangle
=\langle \sin \phi(t,\vec{r})\rangle-h_{2}(s,\vec{r}')
\langle\sin\phi(t,\vec{r})\cos\phi(s,\vec{r}'){\wit\phi(s,\vec{r}')}\rangle
\EEA
It follows that the response function can be expressed as
\BEQ
\rho(t,s;\vec{r},\vec{r}')=\langle\wit{\phi}(s,\vec{r}')
\sin\left(\phi(t,\vec{r})-\phi(s,\vec{r}')\right)\rangle 
\label{eqXYres}
\EEQ
Since both fields $\phi, {\wit \phi}$ are Gaussian, it can further be
shown that (see appendix~A) 
\BEQ
\rho(t,s;\vec{r},\vec{r}')=R(t,s;\vec{r},\vec{r}')\Gamma(t,s;\vec{r},\vec{r}')
\label{eqXYres2}
\EEQ
and we see explicitly that the relationship between $R$ and $\rho$ is
non-trivial. That no higher correlators than the magnetic two-point correlation
function $\Gamma$ enter is a consequence of the Gaussian nature of the 
theory at hand. 

Eqs.~(\ref{eqXYcor},\ref{eqangres},\ref{eqXYres2}) are the main results of
this subsection. Before we can evaluate them, we need some information on the
validity of the spin-wave approximation. 

\subsubsection{Remarks on the validity of the spin-wave approximation}

%%MH cette sous-section est nettement raccourcie
We need a criterion informing us up to what point the results
on $\Gamma$ and $\rho$ derived in the previous subsection within the
spin-wave approximation should be reliable. 

The correlation function $C(t,s;\vec{r},\vec{r}')$ has already been obtained 
above and is given in eq.~(\ref{eqcorfou}). 
For our choice (\ref{eqiniXY}) of initial conditions, its Fourier transform 
$\wht{C}(t,s;\vec{q})$ is \cite{Cugl94b,Rute95a} 
\BEQ
\wht{C}(t,s;\vec{q})=(T_{i}-T_{f})\wht{G}(t+s;\vec{q})
+T_{f}\wht{G}(t-s;\vec{q})
\label{eqcorfin}
\EEQ
where $\wht{G}$ is given by
\BEQ
\wht{G}(u;\vec{q}) :=
\frac{1}{{\vec{q}}^{2}}\exp\left(-{\vec{q}}^{2}(u+\Lambda^2) \right)
\label{eqcorfin2}
\EEQ
and we have explicitly introduced an {\sc uv}-cutoff which simulates the 
lattice spacing (we shall let $\Lambda\to 0$ at the end). Therefore, 
a two-point correlation function $\langle \phi\phi\rangle$ is of order
${\rm O}(T_i,T_f)$ whereas a response function $\langle\phi\wit{\phi}\rangle$
is of order ${\rm O}(1)$ in the initial and final temperatures. 

In order to discuss further the validity of the spin-wave approximation, 
we keep the next-order term as well and consider the Hamiltonian
\BEQ
{\cal H[\phi]}= {\cal H}_{0}+\frac{1}{2}\int \!\D\vec{r} \:
\left[ \left(\nabla \phi(\vec{r})\right)^{2}+g_{4}
\left( \nabla\phi(\vec{r}) \right)^{4}\right] 
\EEQ
where $g_4$ is some constant. A straightforward calculation shows that to
first order in $g_4$, the correction to the spin-wave approximation of 
the two-point correlation function is given by 
\BEQ \label{eqpertur}
\delta C(t,s;\vec{r},\vec{r}') = \int \!\D u\,\D\vec{R}\: 
\left\langle\phi(t,\vec{r})\phi(s,\vec{r}'){\wit \phi}(u,\vec{R})
\nabla\left( \nabla\phi(u,\vec{R}) \right)^{3}\right\rangle
\simeq {\rm O} 
\left( \left( \frac{T_{i}}{T_{c}(d)} \right)^{2},
\left( \frac{T_{f}}{T_{c}(d)} \right)^{2} \right)
\EEQ
where the six-point function is factorized into two-point function by
Wick's theorem. 
As a result, {\it the spin-wave approximation is a first-order 
approximation in the initial and final temperatures}. 
Consistent expressions of two-point
functions must be expanded to first order in $T_i,T_f$. Higher-order terms
in $T_{i,f}$ calculated within the spin-wave approximation should not be
expected to be reliable. 

\subsubsection{Correlation and response functions in the spin-wave 
approximation}

Finally we are ready to list the result for two-time correlation and
response functions in the spin-wave approximation. From the previous
subsection we know that $C$ must be of first order in $T_i,T_f$. The consistent
result for the magnetic correlation function is therefore obtained by
expanding eq.~(\ref{eqXYcor}) to first order in 
temperature. Then 
\BEQ
\Gamma(t,s;\vec{r},\vec{r}')\simeq 1+C(t,s;\vec{r},\vec{r}')-\frac{1}{2}
\left[ C(t,t;\vec{r},\vec{r})+C(s,s;\vec{r}',\vec{r}')\right]
\label{eqXYswacor}
\EEQ
A more suggestive form of this is found as follows. We have 
\BEA
\langle \vec{S}(t,\vec{r}) \rangle \cdot\langle \vec{S}(s,\vec{r}')\rangle
&=&\langle \cos\phi(t,\vec{r})\rangle \langle \cos\phi(s,\vec{r}')\rangle
+\langle \sin\phi(t,\vec{r})\rangle \langle \sin\phi(s,\vec{r}')\rangle
\nonumber \\
&\simeq& 1 - \frac{1}{2} \langle\phi(t,\vec{r})^2\rangle 
\langle\phi(s,\vec{r}')^2\rangle + \langle\phi(t,\vec{r})\rangle
\langle\phi(s,\vec{r}')\rangle +\cdots 
\EEA
where in the second line we performed a low-temperature expansion which
must be kept to second order in $\phi$ in order to be of first order in
the temperature, since $C=\langle\phi\phi\rangle={\rm O}(T)$. 
Furthermore, because of the $\phi\mapsto -\phi$ inversion
symmetry, $\langle\phi\rangle=0$. Inserting this into (\ref{eqXYswacor})
we find, of course only in the context of the spin-wave approximation
\BEQ
\langle \vec{S}(t,\vec{r})\cdot\vec{S}(s,\vec{r}') \rangle
-\langle \vec{S}(t,\vec{r}) \rangle \cdot\langle \vec{S}(s,\vec{r}')\rangle
=C(t,s;\vec{r},\vec{r}')
\label{eqXYcorcon}
\EEQ
and the relation between $\Gamma$ and $C$ is finally clarified (the equilibrium 
version of this is well-known, see \cite{Itzy89}, sect. 4.2.2).

For notational simplicity, we shall now concentrate on the autocorrelation
and autoresponse functions. First, the angular correlation function
is given by eqs.~(\ref{eqcorfin}) and (\ref{eqcorfin2}). We have
\BEQ \label{eqexprcor}
\renewcommand{\arraystretch}{2.0}
C(t,s)= \left\{ {{ \begin{array}{*{20}ll} 
\frac{2(4\pi)^{-d/2}}{d-2}\left[ (T_{i}-T_{f})(t+s+\Lambda^2)^{1-d/2}
+T_{f}(t-s+\Lambda^2)^{1-d/2} \right] & & \mbox{\rm ~~;~ if $d>2$} \\
(4\pi)^{-1}\left[(T_{f}-T_{i})\ln(t+s+\Lambda^2)
-T_{f}\ln(t-s+\Lambda^2)\right]  & &
\mbox{\rm ~~;~ if $d=2$}.
\end{array} }} \right. 
\renewcommand{\arraystretch}{1.0}
\EEQ 
Second, the autoresponse functions are given by 
\BEA
\rho(t,s) &=& R(t,s)\left(1+C(t,s)-\frac{1}{2}
\left[ C(t,t)+C(s,s)\right]\right)
\nonumber \\
R(t,s) &=& \left[4\pi(t-s+\Lambda^2) \right] ^{-d/2}
\label{eqrho1}
\EEA
These results require a detailed discussion.
\begin{enumerate}
\item The $2D$ XY model was studied in detail in the spin-wave approximation
before \cite{Rute95a,Bert01} and we now show that their results, although 
they might at first sight appear to be different, agree with 
eqs.~(\ref{eqXYswacor},\ref{eqexprcor},\ref{eqrho1}). For notational
simplicity, we restrict to $T_i=0$, $T_f=T$. First, the magnetic
autocorrelation function is \cite[eq.~(11)]{Bert01} 
\BEA
\lefteqn{ \Gamma(t,s) = 
\langle \vec{S}(t,\vec{r})\cdot\vec{S}(s,\vec{r})\rangle
\:=\: \left( \frac{\Lambda^4(t+s+\Lambda^2)^2}
{(2t+\Lambda^2)(2s+\Lambda^2)(t-s+\Lambda^2)^2}\right)^{\eta(T)/4} 
}
\nonumber \\
&=& \exp\left[ \frac{\eta(T)}{4} \ln\left( \frac{\Lambda^4 (t+s+\Lambda^2)^2}
{(2t+\Lambda^2)(2s+\Lambda^2)(t-s+\Lambda^2)^2}\right) \right]
\\
&\simeq& 1+ \frac{T}{4\pi}\left( \ln(t+s+\Lambda^2)-\ln(t-s+\Lambda^2)
-\frac{1}{2}\ln(2t+\Lambda^2)-\frac{1}{2}\ln(2s+\Lambda^2)+\ln\Lambda^2\right)
+ {\rm O}(T^2) \nonumber 
\EEA
since the spin-wave approximation is only consistent to lowest order in $T$. 
It is now clear that the above result is reproduced by inserting $C(t,s)$
from eq.~(\ref{eqexprcor}) with $d=2$ into (\ref{eqXYswacor}). Second, the
linear spin response in $2D$ is \cite[eq.~(13)]{Bert01}
\BEA
\rho(t,s) &=& \lim_{\vec{h}\to 0}\frac{\delta\langle \vec{S}(t,\vec{r})\rangle}
{\delta\vec{h}(s,\vec{r})} \:=\: 
\frac{1}{4\pi (t-s+\Lambda^2)} 
\left( \frac{\Lambda^4(t+s+\Lambda^2)^2}
{(2t+\Lambda^2)(2s+\Lambda^2)(t-s+\Lambda^2)^2}\right)^{\eta(T)/4} 
\nonumber \\
&=& \frac{1}{4\pi (t-s+\Lambda^2)}\, \Gamma(t,s)
\EEA
in agreement with eq.~(\ref{eqrho1}) with $d=2$, as it should be.

In $2D$, the results for $\Gamma(t,s)$ and $\rho(t,s)$ were confirmed by
a simulational study with an ordered initial state and $T_f<T_c$ \cite{Abri04}. 

\item For $T_{i}=T_{f}$, the two-point functions are stationary. This is 
only to be expected, since in this case the system was prepared in an 
equilibrium state and remains there. 
\item Ageing occurs when $T_{i}\neq T_{f}$. Time-translation invariance is
broken and we proceed to analyse the resulting scaling behaviour. Now, there
are in principle two equally appealing sets of variables. First, we may choose
to work with the angular correlation function $C$ and its associated response
$R$. Recalling the scaling forms introduced in section~1 
(see eqs.~(\ref{1:gl:SkalCR},\ref{1:gl:lambdaCR})) we shall characterize them
by the exponents $a,b,\lambda_C,\lambda_R$. Second, we may prefer instead 
to work with the magnetic correlation function $\Gamma$ and its associated
response $\rho$. We shall use the same scaling forms, but for clarity we shall
denote the corresponding exponents by $a',b',\lambda_C',\lambda_R'$. 
These exponents are straightforwardly read off in the ageing regime where
$t,s,t-s\gg \Lambda$ and we collect the results in table~\ref{tab1}. 
The exponent $\lambda_C'=(\eta_i+\eta_f)/2$ was already known \cite{Rute95a}. 

%%++++++++++++++++++++++++++++++++++++++++++++++++++++++++++++++++++++++++++++++
\begin{table}[ht]
\caption{Ageing exponents of the $d$-dimensional XY model in the spin-wave
approximation. Here $\eta_{i,f}=\eta(T_{i,f})=T_{i,f}/(2\pi)$ describe the 
initial and  final correlation exponents. \label{tab1}}
\begin{center}
\begin{tabular}{|c|cccc|}  \hline
\multicolumn{5}{|c|}{angular correlation and response} \\ \hline
      & $a$     & $b$   & $\lambda_C$ & $\lambda_R$    \\ \hline
$d=2$ & 0       & 0     & 0           & 2              \\
$d>2$ & $d/2-1$ & $d/2-1$ & $d$       & $d$            \\ \hline
\multicolumn{5}{|c|}{magnetic correlation and response}              \\ \hline
      & $a'$       & $b'$       & $\lambda_C'$        & $\lambda_R'$ \\ \hline
$d=2$ & $\eta_f/2$ & $\eta_f/2$ & $(\eta_i+\eta_f)/2$ & $2+(\eta_i+\eta_f)/2$ \\
$d>2$ & $d/2-1$    & $d/2-1$    & $d$                 & $d$          \\ \hline
\end{tabular} \end{center} 
\end{table}
%%++++++++++++++++++++++++++++++++++++++++++++++++++++++++++++++++++++++++++++++

We see that the exponents satisfy the equalities $a=b$ and $a'=b'$ expected
for nonequilibrium critical dynamics and point out that in $2D$, the 
autocorrelation and autoresponse exponents are different:
$\lambda_R-\lambda_C=\lambda_R'-\lambda_C'=2$. This effect comes from 
the non-disordered initial condition of the spins, as explained first 
in \cite{Pico02}. 
\item Having discussed the values of the ageing exponents, we wish to 
compare the form of the scaling functions with the predictions of local 
scale-invariance as derived in section 4. This requires, however, the correct 
identification of the quasiprimary fields in our system, see section~2. 
It is {\em only}
the quasiprimary fields which are expected to transform in a simple way
under a local scale-transformation and the transformation laws of more
complicated fields built from quasiprimary fields must derived accordingly. 

In the case at hand, it is clear from the complicated structure of the 
magnetic correlation and response functions that the magnetic
order parameter $\vec{S}(t,\vec{r})$ does {\em not} correspond to a quasiprimary
field. Rather, the quasiprimary field should be identified with the phase
$\phi(t,\vec{r})$. Indeed, the form of the angular response $R(t,s)$ is 
in perfect agreement with the prediction (\ref{eqR5},\ref{4:gl:k}) of local 
scale-invariance. This suggests that the response field $\wit{\phi}(s,\vec{r})$
should be quasiprimary as well. 
\item Having thus identified $\phi$ and $\wit{\phi}$ as quasiprimary
fields of the model, it is now clear that the angular autocorrelation function
$C(t,s)$ should be compared to the critical dynamics correlation function as
derived in section 4. However, a direct comparison with eq.~(\ref{eqC16}) is
not possible, since in its derivation fully disordered initial conditions
were assumed. 

We shall therefore proceed in two steps. First, we shall consider the case
$T_i=0$. Because of our initial conditions (\ref{eqiniXY}) the initial 
correlator then vanishes and eq.~(\ref{eqC16}) should now hold
true. Second, we shall show that in the context of the free-field theory
underlying the spin-wave approximation of the XY model, the restriction
to uncorrelated initial states can be lifted. 

We now set $T_i=0$ and $T_f=T$. From the exponents in table~\ref{tab1}, the
predicted autocorrelator scaling function follows from (\ref{eqC16}) as
\BEQ
f_C(y) = \Phi_0 \int_0^1\!\D\theta\, \left( y+1-2\theta\right)^{-d/2}
\;\; ; \;\; d\geq 2
\EEQ
and we see immediately that this is in agreement with the explicit 
angular correlator (\ref{eqexprcor}), upon identification of $\Phi_0$. 

Finally, if we also allow for $T_i>0$, there is a contribution to $C(t,s)$
from the initial condition. We then return to the basic result (\ref{eqC4})
and decompose $C(t,s)=C_{th}+C_{pr}$. The thermal term $C_{th}$ was treated
before and the preparation term is analysed in appendix~B, with the result
\BEQ
C_{pr} = \frac{2(4\pi)^{-d/2}}{d-2} T_i (t+s)^{1-d/2} \;\; ; \;\; d>2
\EEQ
in complete agreement with (\ref{eqexprcor}). The case $d=2$ is treated
similarly.
\end{enumerate}

In conclusion, the two-time autocorrelation and
autoresponse functions of the XY model treated in spin-wave approximation
are in perfect agreement with local scale-invariance, {\em provided} the
angular variable $\phi$ and its associated response field are identified
as the quasiprimary fields of the model. 

\subsection{Fluctuation-dissipation relations in the XY model}

Having checked that both correlation and response functions agree with the
local scale-invariance prediction, we now inquire what can be said on the
approach of the model towards equilibrium. A convenient way to study this
is through the so-called fluctuation-dissipation ratio \cite{Cugl94b}, 
see section~1. Since we have seen that in the XY model angular and magnetic 
observables behave quite differently, it is convenient to define two distinct 
fluctuation-dissipation ratios, namely  
\BEA
\Xi(t,s) &:=& T_f\rho(t,s)
\left(\frac{\partial \Gamma(t,s)}{\partial s}\right)^{-1}
\nonumber \\
X(t,s) &:=& T_fR(t,s)\left(\frac{\partial C(t,s)}{\partial s}\right)^{-1}
\label{eqX2}
\EEA 
Of particular interest will be the limit fluctuation-dissipation
ratios $X_{\infty}$ defined in eq.~(\ref{1:gl:FDR}) and similarly 
$\Xi_{\infty}$. We have seen before that the 
scaling of autocorrelation and autoresponse 
functions is according to the expectations of nonequilibrium critical
dynamics. In this case, according to the Godr\`eche-Luck conjecture 
\cite{Godr00b}, $X_{\infty}$ and $\Xi_{\infty}$ should be universal.

We shall use the available exact results in the XY model to test this 
conjecture by studying the dependence of $X$ and $\Xi$ on the ratio 
$\alpha:=T_i/T_f$ of initial and final temperatures.

\subsubsection{Fluctuation-dissipation ratio for magnetic variables} 

The fluctuation-dissipation ratio $\Xi(t,s)$ obtained from
the magnetic correlation and response functions reads, with $y=t/s$
\BEQ
\frac{1}{\Xi(y)}
=1+\left(1-\frac{T_i}{T_f}\right)\left[\left(\frac{y-1}{y+1}\right)^{d/2}
-\left(\frac{y-1}{2}\right)^{d/2}\right]
\label{eqX3}
\EEQ
For $T_i=0$ and $d=2$ this was already known \cite{Bert01}. 
In the quasiequilibrium regime $y\simeq1$ the fluctuation-dissipation
theorem should hold. Indeed we find $\lim_{y\to 1}\Xi(y)=1$ which confirms 
that $\Xi$ should be well-defined. For large values of $y$, that is for 
well-separated times, we have
\BEQ
\Xi(y)\simeq\frac{T_f}{T_i-T_f}\left(\frac{y-1}{2}\right)^{-d/2}
\label{eqX4}
\EEQ 
and therefore $\Xi_{\infty}=0$, indeed a universal constant. 
It is remarkable that the asymptotic value of $\Xi(y)$ should be independent 
of $d$ and that is agrees with the value $\Xi_{\infty}=0$ of phase-ordering 
kinetics in the low-temperature phase with an ordered, non-critical equilibrium 
state. This kind of result should be more typical of an ordered 
ferromagnetic equilibrium state as it occurs for $d>2$ but is not really
expected for $d=2$ since the equilibrium $2D$ XY model is critical even below 
$T_{\rm KT}$. 

Finally, for large $y$ the asymptotic form of $\Xi(y)$ is independent on
whether the system is cooled or heated. The temperatures merely enter into
a scaling amplitude. 
%%MH Finally, for large $y$ only the sign of $\Xi(y)$ but not its absolute 
%%MH value depends on whether the system is cooled or heated.  
            
\subsubsection{Fluctuation-dissipation ratio for angular variables}

In the same way, the fluctuation-dissipation ratio for the angular
variables is found. It reads
\BEQ
\frac{1}{X(y)}=1+\left(1-\frac{T_i}{T_f}\right)
\left(\frac{y+1}{y-1}\right)^{-d/2}
\label{eqX5}
\EEQ
As before, in the quasiequilibrium regime $t\simeq s$, $X(t,s)=1$. Surprisingly,
however, for large values of $y=t/s$, the limit fluctuation-dissipation ratio 
\BEQ
X_{\infty}=\left(2-\frac{T_i}{T_f}\right)^{-1}
\label{eqX6}
\EEQ 
depends continuously on $\alpha=T_i/T_f$. We recall from table~\ref{tab1} that 
the non-equilibrium exponents of the {\em angular} variables are all independent
of both $T_{i}$ and $T_f$ and although the exponents do depend on $d$, 
we see from (\ref{eqX6}) that $X_{\infty}$ does not. Taken literally, 
this would be an example of a non-universal value of the limit
fluctuation-dissipation ratio. 

We recall that most `physically reasonable'
systems undergoing nonequilibrium critical dynamics one usually finds
$0\leq X_{\infty}\leq 1/2$, see \cite{Godr02} for a review. Motivated
from mean-field theories of spin glasses, it is sometimes suggested that
$T_{\rm eff}:= T/X_{\infty}$ might be interpreted as an effective temperature 
for which the fluctuation-dissipation theorem would hold. It is hard to see
how in this case ($X_{\infty}$ may even become negative) 
such an interpretation could be maintained.

%%%%%%%%%%%%%%%%%%%%%%%%%%%%%%%%%%%%%%%%%%%%%%%%%%%%%%%%%%%%%%%%%%%%%%%%%%%%%%%%
\section{The critical voter model in $d$ dimensions}
%%%%%%%%%%%%%%%%%%%%%%%%%%%%%%%%%%%%%%%%%%%%%%%%%%%%%%%%%%%%%%%%%%%%%%%%%%%%%%%%

We now study a qualitatively different type of application of local 
scale-invariance in the so-called voter model, see \cite{Ligg99} and references
therein. The model describes the temporal evolution of configurations $\cal C$ 
of spins $\sigma_{\vec{r}}(t)=\pm 1$ on a $d$-dimensional hypercubic lattice 
$\mathbb{Z}^d$. The dynamics is assumed to be given by a master equation
\BEQ 
\frac{\D}{\D t}P({\cal C};t) = \sum_{\vec{r}\in\mathbb{Z}^d} 
\left[ W_{\vec{r}}({\cal C}^{(\vec{r})}) P({\cal C}^{(\vec{r})};t) - 
W_{\vec{r}}({\cal C}) P({\cal C};t) \right]
\EEQ
Here the configuration ${\cal C}^{(\vec{r})}$ is obtained from $\cal C$
by inverting the single spin at site $\vec{r}$. Finally, the transition
rates for a spin reversal $\sigma_{\vec{r}}\mapsto -\sigma_{\vec{r}}$ are
given by
\BEQ \label{6:gl:VRaten}
W_{\vec{r}}({\cal C}) = \frac{1}{2} \left[ 1 - \frac{1}{2d} \sum_{k=1}^d
\left( \sigma_{\vec{r}}\sigma_{\vec{r}+\vec{e}_k} 
+ \sigma_{\vec{r}}\sigma_{\vec{r}-\vec{e}_k} \right) \right]
\EEQ
where the $\vec{e}_k$, $k=1,\ldots d$ form an orthonormalized basis of unit
vectors on the $d$-dimensional hypercubic lattice. 

With respect to the kinetic spherical model and the XY model studied
previously, the voter model is different since in general it does {\em not} 
satisfy detailed balance and therefore will not relax to an equilibrium state. 
By considering a general kinetic Ising model with a dynamics respecting the
global $\mathbb{Z}_2$-symmetry, it can be shown that the
transition rates (\ref{6:gl:VRaten}) correspond to the critical point of the
so-called linear voter model \cite{Oliv93,Sast03}. 
The non-equilibrium kinetics of
the critical voter model (\ref{6:gl:VRaten}) has been studied in detail by
Dornic \cite{Dorn02}. In $d=1$ dimensions the model
co\"{\i}ncides with the kinetic Glauber-Ising model at zero temperature (which
we shall revisit in appendix~C) but for $d>1$ these two models are different. 
In particular, it is known that the domain growth of the voter model is not 
driven by the minimization of the surface tension between the two 
phases \cite{Dorn01,Dorn02} but which is the mechanism which drives ageing
in simple ferromagnets \cite{Bray94,Bray00}. 

We are interested here in the correlation functions 
$C_{\vec{r}}(t)=\langle \sigma_{\vec{r}}(t)\sigma_{\vec{0}}(t)\rangle$ and
$C_{\vec{r}}(t,s)=\langle \sigma_{\vec{r}}(t)\sigma_{\vec{0}}(s)\rangle$ which
are easily seen to satisfy the following equations of 
motion \cite{Frac96,Dorn02}
\BEA
\frac{\partial}{\partial t}C_{\vec{r}}(t) &=& 2\Delta_{\vec{r}}C_{\vec{r}}(t)
\nonumber \\
\frac{\partial}{\partial t}C_{\vec{r}}(t,s) &=& \Delta_{\vec{r}}C_{\vec{r}}(t,s)
\EEA
subject to the following boundary conditions
\BEQ
C_{\vec{0}}(t)=1 \;\; , \;\; C_{\vec{r}}(0)=\delta_{\vec{r},\vec{0}} \;\; ,\;\;
C_{\vec{r}}(t,t) = C_{\vec{r}}(t)
\EEQ
where $\Delta_{\vec{r}}$ is the discrete Laplacian and where the initial 
magnetization $\sum_{\vec{r}} \langle \sigma_{\vec{r}}(0)\rangle=0$. For the 
autocorrelation function $C(t,s)=C_{\vec{0}}(t,s)$ of the critical voter model 
(\ref{6:gl:VRaten}) one finds in the ageing regime, where
$t,s$ and $t-s$ are all sufficiently large \cite{Dorn02}, with $y=t/s$
\BEQ \label{eqv2}
\renewcommand{\arraystretch}{2.0}
C(t,s) = \left\{ {{ \begin{array}{*{20}ll} 
\frac{2}{\pi}\arctan\sqrt{{2}/{(y-1)}\:} & & \mbox{\rm ~~;~ if $d=1$} \\
\ln(s)^{-1}\,\ln\left((y+1)/(y-1)\right) & & \mbox{\rm ~~;~ if $d=2$} \\
s^{-d/2+1}\left(\frac{d}{2\pi}\right)^{d/2}
\frac{2\gamma_{d}}{d-2}\left[\left(y-1\right)^{-d/2+1}
-\left(y+1\right)^{-d/2+1}\right] & & \mbox{\rm ~~;~ if $2< d< 4$} 
\end{array} }} \right.
\renewcommand{\arraystretch}{1.0}
\EEQ
where $\gamma_{d}$ is the probability that a random walk in $d$ dimensions 
and starting from the origin never returns. We are not aware of published
results on $R(t,s)$ for $2<d<4$ in this model. Clearly, time-trans\-la\-tion
invariance is broken for all $d\geq 2$.    

We wish to compare (\ref{eqv2}) with the prediction eq.~(\ref{eqC16}) of local 
scale-invariance. The case $d=1$ will be dealt with in appendix~C and since for
$d=2$ logarithmic scaling is found, the form of local scale-invariance as
presented here is not applicable.\footnote{An extension of Schr\"odinger
invariance to logarithmic Schr\"odinger invariance in analogy to logarithmic
conformal invariance, see e.g. \cite{Henk99}, might be needed here.} 
We therefore concentrate on the dimensions $2<d<4$. From the asymptotics of
$C(t,s)$ in (\ref{eqv2}) we read off the exponents 
\BEQ
b = \frac{d}{2} -1 \;\; , \;\; \lambda_C = d 
\EEQ
since the exponent $z=2$ is known, see \cite{Ligg99,Dorn02}. 
{}From eq.~(\ref{eqC16}) we expect
\BEQ
f_{C}(y) = \Phi_{0,c} \int_{0}^{1} \!\D\theta\,\left( y+1-2\theta\right)^{-d/2}
= \frac{\Phi_{0,c}}{d-2} \left[ (y-1)^{-d/2+1} - (y+1)^{-d/2+1}\right]
\EEQ
in full agreement with (\ref{eqv2}) and we identify the normalization
constant $\Phi_{0,c}=2\gamma_{d}\left({d}/{2\pi}\right)^{d/2}$. Indeed, we 
see that the form (\ref{Phi0c}) -- which in principle is only valid 
asymptotically -- is in fact exact in the critical voter model. This is
not surprising in view of the underlying free-field theory. 

In conclusion, for the critical voter model with two competing steady states
ageing occurs. The scaling form of the two-time autocorrelation function is
in exact agreement with the prediction of local scale-invariance. This is the
first time such an agreement is found for a system without detailed balance. 

%%%%%%%%%%%%%%%%%%%%%%%%%%%%%%%%%%%%%%%%%%%%%%%%%%%%%%%%%%%%%%%%%%%%%%%%%%%%%%%%
\section{The free random walk}
%%%%%%%%%%%%%%%%%%%%%%%%%%%%%%%%%%%%%%%%%%%%%%%%%%%%%%%%%%%%%%%%%%%%%%%%%%%%%%%%

%%MH la dependance en r est supprimee

Last, but not least, we briefly consider the simplest example of a
system undergoing ageing: the free random walk \cite{Cugl94b}. The 
Langevin equation describing the time-evolution of the order parameter
$\phi$ reads 
\BEQ \label{eqRW1}
\frac{\partial \phi(t)}{\partial t}=h(t)+\eta(t)
\EEQ   
where a deterministic external field $h(t)$ has been added in order to
be able to compute response functions. The Gaussian noise $\eta$ is
characterized as usual by its first two moments, see eq.~(\ref{eqeta}).
Here, we choose the notation such that the relationship to local 
scale-invariance becomes evident. 

The autocorrelation and linear autoresponse
functions were already calculated by Cugliandolo et {\it al.} \cite{Cugl94b}.
They obtained, with the initial condition $C(t,0)=0$   
\BEA
R(t,s) &=& \left.\frac{\delta\langle\phi(t)\rangle}{\delta h(s)}\right|_{h=0}
\:=\: \frac{1}{2T} \langle\phi(t)\eta(s)\rangle \:=\: \Theta(t-s)
\nonumber \\
C(t,s) &=& \langle \phi(t)\phi(s)\rangle \:=\: 2T\min(t,s) 
\label{eqRW2}
\EEA 
Clearly, the system undergoes ageing, since $C(t,s)$ does not merely depend
on $t-s$ and furthermore, it stays forever out of equilibrium, since the
fluctuation-dissipation ratio $X(t,s)=1/2$ \cite{Cugl94b}. 

We wish to check that the results (\ref{eqRW2}) are compatible with the
predictions of local scale invariance. For the autoresponse function,
the first eq.~(\ref{eqRW2}) is clearly compatible with
(\ref{eqR4},\ref{4:gl:k},\ref{4:gl:xdg}),
with the exponents $a=-1$ and $\lambda_R=0$. For the autocorrelation
function, we expect from (\ref{eqC5})
\BEQ \label{7:gl:C}
C(t,s) = T \int \!\D u\;  
\left.\frac{\delta^2\langle\phi(t)\phi(s,)\rangle}{\delta h(u)^2}\right|_{h=0}
\EEQ
where in view of the initial condition $C(t,0)=0$ used in eq.~(\ref{eqRW2}) we
assumed a vanishing initial correlator. 
  
In order to calculate the above derivative, we solve the Langevin equation for
a given field $h$ and obtain the autocorrelation function 
\BEQ
C(t,s;[h])=C(t,s;[0])+\int_{0}^{s}\int_{0}^{t}\!\D v\D w\: h(v)h(w)
\label{eqRW3}
\EEQ  
where $C(t,s;[0])=C(t,s)$ is of course given by the second eq.~(\ref{eqRW2}). 
The required second derivative of the autocorrelation function becomes
in a field-theoretic formulation some three-point correlator 
(see eq.~(\ref{eqR03})) and it is now easy to see that 
\BEQ
R_{0}^{(3)}(t,s,u)=
\left.\frac{\delta^{2}C(t,s;[h])}{\delta h(u)^2}\right|_{h=0}
=2\Theta(t-u)\Theta(s-u)
\label{eqRW4}
\EEQ 
Inserting this into (\ref{7:gl:C}), the desired result for $C(t,s)$ in
eq.~(\ref{eqRW2}) is indeed recovered. We read off the exponents $b=-1$ and
$\lambda_C=0$. Of course, the exponent equalities $a=b$ and 
$\lambda_C=\lambda_R$ are a necessary requirement for having a non-vanishing
limit fluctuation-dissipation ratio $X_{\infty}=1/2$. 

In conclusion, the evidence from the two-time autocorrelation and autoresponse
function of the free random walk is fully consistent with 
local scale-invariance.

%%%%%%%%%%%%%%%%%%%%%%%%%%%%%%%%%%%%%%%%%%%%%%%%%%%%%%%%%%%%%%%%%%%%%%%%%%%%%%%%
\section{Conclusions and discussion}
%%%%%%%%%%%%%%%%%%%%%%%%%%%%%%%%%%%%%%%%%%%%%%%%%%%%%%%%%%%%%%%%%%%%%%%%%%%%%%%%

We have analysed the ageing behaviour in systems with a non-conserved 
order parameter and described by a Langevin equation. Our main assumption
was that the noiseless part of that Langevin equation is Galilei-invariant.
Together with dynamical scaling this hypothesis fixes the dynamical exponent 
$z=2$ and implies for local theories Schr\"odinger-invariance \cite{Henk03}. 
There are good reasons for admitting such a hypothesis. For example the 
phase-ordering kinetics of the Glauber-Ising model in $d>1$ dimensions quenched 
to a temperature $T<T_c$ provides strong evidence that the scaling function 
of its space-time response function $R(t,s;\vec{r},\vec{r}')$ has the form
predicted from Galilei-invariance \cite{Henk03b}. However, since groups of
local scale transformations such as the Schr\"odinger group are dynamical 
symmetries of noiseless differential equations only, the r\^ole of the noise
in the Langevin equation or from the initial conditions has to be addressed. 

We have carried out such a study, for the important special case where
$z=2$ and the initial state is fully disordered. Considering the Langevin
equation as the classical equation of motion of a MSR-type field theory, we
have calculated the two-time correlation and linear response functions by
studying how the dynamical symmetry properties of the noiseless part of that
field theory are reflected in these noisy averages. These averages can be
written in form of a perturbative expansion around the noiseless theory
and we have shown that only a {\em finite} number of terms in these series
contributes. Specifically, we have found:
\begin{enumerate}
\item The two-time linear response function $R=\langle\phi\wit{\phi}\rangle$ 
involving only quasiprimary fields
is independent of both the thermal and the initial noise. This explains
why the form (\ref{1:gl:fR}) of scaling function $f_R(y)$ of the linear
response -- previously derived from the symmetries of the noiseless theory -- 
has been reproduced in many different systems with $T>0$ either exactly or 
with a considerable numerical precision 
\cite{Cala02a,Cann01,Cugl94b,Godr00b,Godr02,Henk01,Henk02,Henk02a,Henk03b,Jans89,Newm90,Pico02}.
\item We obtain the reduction formula eq.~(\ref{eqC4}) which expresses
the two-time correlation function in terms of certain noiseless three- and
four-point response functions. For the uncorrelated initial conditions
(\ref{4:gl:aini}) only a single noiseless three-point response function is
needed.
\item The scaling forms of correlation and response functions are governed by 
the two non-trivial exponents $\lambda_C$ and $\lambda_R$ which are in general 
distinct from each other. Given the initial correlator (\ref{4:gl:aini}), local 
scale-invariance with $z=2$ provides a sufficient condition for the exponent 
equality
\BEQ
\lambda_C = \lambda_R
\EEQ
\item The scaling of the two-time autocorrelator $C(t,s)$ is described by a 
scaling function $\Phi(w)$ which in turn depends on a scaling function 
$\Psi(\rho)$ which arises in the three-point function of quasiprimary fields
in Schr\"odinger-invariant theories. Depending on whether the thermal or the
initial noise is dominant, two distinct scaling forms
eqs.~(\ref{eqC11}) and (\ref{eqC15}) are found and we have argued that they 
should describe the cases when the system is quenched to temperatures $T<T_c$ 
and $T=T_c$, respectively (in the latter case only if $z=2$ still holds after
renormalization). 
\item Schr\"odinger-invariance by itself does not determine the form of 
$\Psi(\rho)$. We have argued that the related three-point response function 
should be non-singular and then find the asymptotic behaviour for $w\to\infty$
\BEQ \label{8:gl:2}
\Phi(w) \sim  w^{-\vph} \;\; , \;\; \vph = \left\{
\begin{array}{ll} d/2 - \lambda_C & \mbox{\rm ~~;~ if $T<T_c$} \\
                  1+a             & \mbox{\rm ~~;~ if $T=T_c$} 
\end{array} \right.
\EEQ
This suggests the following approximate forms, with $y=t/s$
\BEQ \label{8:gl:3}
\renewcommand{\arraystretch}{2.0}
C(t,s) \approx \left\{ \begin{array}{ll}
M_{\rm eq}^2 \left[ (y+1)^2/(4y)\right]^{-\lambda_C/2} 
  & \mbox{\rm ~~;~ if $T<T_c$} \\
\Phi_{0,c}\, y^{1+a-\lambda_C/2} \int_0^1 \!\D\theta\, 
\theta^{\lambda_C-2-2a} (y+1-2\theta)^{-1-a} 
  & \mbox{\rm ~~;~ if $T=T_c$}
\end{array} \right.
\renewcommand{\arraystretch}{1.0}
\EEQ
At least, these forms are consistent with the required asymptotic behaviour of
$f_C(y)$ as $y\to\infty$, see section~1. For free-field theories (\ref{8:gl:2}) 
holds for all values of $w$ and then (\ref{8:gl:3}) becomes exact. 

In the past, approximate expressions for the scaling of magnetic 
correlation functions were derived from Gaussian closure procedures for 
kinetic O($n$)-models undergoing phase-ordering kinetics, 
see \cite{Bray94}. This gives for the magnetic 
autocorrelation function at $T=0$ \cite{Bray91a,Bray92,Roja99}
\BEQ \label{8:gl:4}
f_{C,{\rm BPT}}(y) \approx \frac{n}{2\pi} 
\left[ B\left(\frac{1}{2},\frac{n+1}{2}\right)\right]^2 
\left(\frac{4y}{(y+1)^2}\right)^{d/4}
{_2F_1}\left(\frac{1}{2},\frac{1}{2};\frac{n+2}{2};
\left(\frac{4y}{(y+1)^2}\right)^{d/2}\right) 
\EEQ
where $B$ is Euler's beta function and ${_2F_1}$ a hypergeometric function.
It is easy to see that the scaling function (\ref{8:gl:3}) with $T=0$ is
recovered from (\ref{8:gl:4}) in the $n\to\infty$ limit. 
In the notation of section~5.2, eq.~(\ref{8:gl:4}) implies an 
exponent $\lambda_C'=d/2$. This is a typical value
for a free-field theory for $T<T_c$ (which indeed describes the O($n$)-model
in the $n\to\infty$ limit) but which will in general not hold for $n<\infty$ 
and thus (\ref{8:gl:4}) cannot be expected to represent well the behaviour for 
$y=t/s$ large.\footnote{Indeed, to leading order in $1/n$ and for uncorrelated
initial conditions, 
$\lambda_C'=d/2+(4/3)^d(d+2)\frac{2d}{9}B(1+d/2,1+d/2) n^{-1}$ \cite{Bray91}.} 
One might wonder whether the long-standing difficulties in 
arriving at scaling functions which cover adequately the whole range of values 
of $y$ should not be related at least partially to
having worked with dynamic variables which might turn out to be not the
most basic ones of the model ?   
\item We have tested these predictions on the exact solutions of the 
kinetic spherical model and the XY model in the spin-wave approximation. In
these cases, the exponent $a$ is given by (\ref{1:gl:a}). In order to
compare the exact model results with the prediction (\ref{8:gl:3}), it was
necessary to carefully identify the quasiprimary fields of the models. For the 
spherical model, the natural magnetic order parameter and its
response field could be used as quasiprimary fields. On the other hand, for 
the XY model it turned out that the coarse-grained magnetic moment is not
quasiprimary but rather the angular variable is. 

These examples underline the importance of the correct identification of the
quasiprimary fields in a given model. 

The explicit results in the XY model also illuminate in a new way the 
Godr\`eche-Luck conjecture on the universality of the limit 
fluctuation-dissipation ratio. Further tests on the form of the autocorrelation
function in the Glauber-Ising model in $d\geq 2$ dimensions are presently
carried out and will be reported elsewhere \cite{Pico03b,Henk04a}. 
\item We also showed that the autocorrelator in the critical voter model
in $2<d<4$ dimensions and which does not satisfy detailed balance, again
agrees with (\ref{8:gl:3}). This is the first example of a new domain of
application of local scale-invariance. We also confirmed (\ref{8:gl:3}) for
the free random walk. 

These four confirmations, although all based on an underlying free-field theory
provide further evidence in favour of a Galilei-invariance of the noiseless 
theory. 
\item The scaling of the linear response of the $1D$ Glauber-Ising model
at $T=0$ can only be explained through a generalization of the representations
of the Lie algebra of local scale-invariance. It would be interesting to
see whether a similar phenomenon could be found in different $1D$ systems
undergoing ageing at $T=0$. 
\end{enumerate} 
Our approach has been based in an essential way on the assumption of
Galilei-invariance of the noiseless theory. However, Mazenko \cite{Maze03} 
recently studied phase-ordering kinetics in the context of the time-dependent
non-linear Ginzburg-Landau equation. He carried out a second-order 
perturbative calculation around a Gaussian theory which is equivalent to the
Ohta-Jasnow-Kawasaki approximation and reports a deviation of the two-time
autoresponse function $R(t,s)$ from the local scale-invariance prediction
(\ref{1:gl:fR}). A similar difficulty had been observed
before by Calabrese and Gambassi who studied non-equilibrium dynamics (that is
$T=T_c$) of the O($n$) model, for both model A and model C dynamics, 
through MSR field theory \cite{Cala02b,Cala02c,Cala03}. Again, already 
their classical action is manifestly not Galilei-invariant.\footnote{At
criticality, the combined effect of thermal and initial fluctuations leads
for interacting theories to a non-trivial value of $z\ne 2$ under
renormalization so that our arguments are no longer directly applicable.} 

At face value, Mazenko's result \cite{Maze03} is in disagreement with the 
simulational data obtained from the $2D$ and $3D$ Glauber-Ising model with 
$T<T_c$ and based on the master equation. These do reproduce 
(\ref{1:gl:fR}) for the autoresponse function $R(t,s)$
\cite{Henk02a,Henk03b,Henk03e} as well as the extension 
for the spatio-temporal response $R(t,s;\vec{r}-\vec{r}')$ \cite{Henk03b}. 
Could this mean that there 
are subtle differences between the formulation of stochastic systems either 
through a master equation or else through a coarse-grained Langevin 
equation\footnote{In $1D$ and at $T=0$, the autocorrelation exponent 
$\lambda_C=1$ found in the Glauber-Ising model \cite{Godr00a,Lipp00} differs
from the exactly known exponent 
$\lambda_C=0.6006\ldots$ \cite{Bray95} determined in the time-dependent 
Landau-Ginzburg equation.} and which affect 
the formal Galilei-invariance of the theory~? Alternatively, if under
renormalization the dynamical exponent $z=2$ remains constant, might 
the theory flow to a fixed point where asymptotically
Galilei-invariance would hold ?

All in all, based on a postulated extension of dynamical scaling to some
local scale-invariance, we have reformulated 
the problem of finding the scaling function
of the two-time autocorrelation function of ageing systems as one of a 
discussion of the properties of certain three-point response functions of the 
noiseless theory. The evidence available at present suggests that this approach 
might be capable of shedding a new light on the issue. Finally, it might
be of interest to search for extensions for dynamical exponents $z\ne 2$
and/or to study ageing systems with a conserved order parameter. Asymptotic
information on the two-point functions in the latter case is now becoming
available \cite{Godr04}.

\newpage

%%%%%%%%%%%%%%%%%%%%%%%%%%%%%%%%%%%%%%%%%%%%%%%%%%%%%%%%%%%%%%%%%%%%%%%%%%%%%%%%
\appsection{A}{On Gaussian integration }
%%%%%%%%%%%%%%%%%%%%%%%%%%%%%%%%%%%%%%%%%%%%%%%%%%%%%%%%%%%%%%%%%%%%%%%%%%%%%%%%

We present the details of the calculations of the magnetic two-time 
correlation function $\Gamma$ and its associated linear response function
$\rho$ for the $d$-dimensional XY model in the spin-wave approximation.
They are defined as  
\BEA
\Gamma(t,s;\vec{r},\vec{r}') &=& \int  \!{\cal D}\phi{\cal D}{\wit \phi}\:  
\cos (\phi(t,\vec{r})-\phi(s,\vec{r}'))  \exp\left(-S[\phi, {\wit \phi}]\right) 
\nonumber \\
\rho(t,s;\vec{r},\vec{r}') &=& \int  \!{\cal D}\phi{\cal D}{\wit \phi}\:  
\wit{\phi}(s,\vec{r}') \sin \left( \phi(t,\vec{r})-\phi(s,\vec{r}') \right)  
\exp\left(-S[\phi, {\wit \phi}]\right)  
\label{eqappC1}
\EEA  
where $S[\phi, {\wit \phi}]$ is the free-field Martin-Siggia-Rose action which
is Gaussian. 

\subsection{The correlation function $\Gamma(t,s;\vec{r},\vec{r}')$}

Using de Moivre's identities
\BEA
\Gamma(t,s;\vec{r},\vec{r}') &=& \frac{1}{2}\left[ \left\langle 
\exp \II(\phi(t,\vec{r})-\phi(s,\vec{r}'))\right\rangle 
+ \left\langle \exp \II(\phi(s,\vec{r}')-\phi(t,\vec{r}))\right\rangle \right] 
\nonumber \\
&=& \frac{1}{2}\left[ \left\langle 
\exp \II\int \!\D u\,\D\vec{R}\: J(u,\vec{R})\phi(u,\vec{R}) \right\rangle 
+ \left\langle \exp -\II\int \!\D u\,\D\vec{R}\: J(u,\vec{R})\phi(u,\vec{R}) 
\right\rangle \right]
\EEA
where $J(u,\vec{R})$ is given by 
\BEQ
J(u,\vec{R})=\delta(u-t)\delta(\vec{R}-\vec{R})
-\delta(u-s)\delta(\vec{R}-\vec{r}')
\label{eqappC2}
\EEQ 
A generally valid result from free-field theory reads, see \cite{Zinn89}  
\BEQ
\left\langle \exp \II\int \!\D u\,\D\vec{R}\: 
J(u,\vec{R})\phi(u,\vec{R}) \right\rangle = 
\exp \left[-\frac{1}{2}\int \!\D u\,\D u'\,\D\vec{R}\,\D\vec{R}' \: 
J(u,\vec{R})\left\langle  \phi(u,\vec{R})\phi(u',\vec{R}')\right\rangle 
J(u',\vec{R}') \right]
\EEQ
which is  one of the various forms of writing  
Wick's theorem. Using the explicit form of the current in (\ref{eqappC2}), 
equation (\ref{eqXYcor}) follows immediately. 

\subsection{The response function $\rho(t,s;\vec{r},\vec{r}')$}

We now focus on the magnetic response function 
$\rho(t,s;\vec{r},\vec{r}')$. We decompose the MSR action 
\BEQ
S[\phi,{\wit \phi},\vec{0}]= S_{0}[\phi,{\wit \phi}]+s[\phi, {\wit \phi}]
\label{eqappC3}
\EEQ 
where 
\BEA
S_{0}[\phi, {\wit \phi}] &=& \int \!\D u\,\D\vec{r}\: {\wit \phi}
\left[\frac{\partial\phi}{\partial u} -\Delta \phi\right]
\nonumber \\
s[\phi, {\wit \phi}] &=& -\int \!\D u\,\D u'\,\D\vec{r}\,\D\vec{r}'\: 
{\wit \phi}(u,\vec{r})\kappa(u,u';\vec{r}-\vec{r}'){\wit \phi}(u',\vec{r}') 
\EEA
and 
\BEQ
\kappa(u,u';\vec{r}-\vec{r})=T\delta(u-u')\delta(\vec{r}-\vec{r}')
+\frac{1}{2}\delta(u)\delta(u')a(\vec{r}-\vec{r}')
\label{eqappC4}
\EEQ
The noiseless two-point functions are 
\BEA
R_{0}(t,s;\vec{r},\vec{r}') 
&=& \langle \phi(t,\vec{r}){\wit \phi}(s,\vec{r}')\rangle_{0} 
= \Theta(t-s)[4\pi(t-s)]^{-d/2}
\exp\left[-\frac{(\vec{r}-\vec{r}')^{2}}{4(t-s)}\right] 
\nonumber \\
C_{0}(t,s;\vec{r},\vec{r}') 
&=& \langle \phi(t,\vec{r})\phi(s,\vec{r}')\rangle_{0}
=0
\nonumber \\
{\wit C}_{0}(t,s;\vec{r},\vec{r}') 
&=& \langle {\wit \phi}(t,\vec{r}){\wit \phi}(s,\vec{r}')\rangle_{0}
=0 
\label{A:B}
\EEA
Below, we shall need the equal-time response function
$\left. R(t,s;\vec{r},\vec{r}')\right|_{t=s}$. To give to this quantity a value,
one may discretize the Langevin equation. This may be done according to
several different schemes, see \cite{Jans92,vanK97}. Here, we shall use the
It\^o prescription which amounts to 
\BEQ
R_{0}(t,t;\vec{r},\vec{r}')=0
\label{eqappC6}
\EEQ   

The magnetic response function reads from (\ref{eqappC1})
\BEQ
\rho(t,s;\vec{r},\vec{r}')=\left\langle{\wit \phi}(s,\vec{r}')
\sin(\phi(t,\vec{r})-\phi(s,\vec{r}'))
\exp -s[\phi, {\wit \phi}] \right\rangle_{0} 
\label{eqappC5}
\EEQ
where $\langle \rangle_0$ is the average with the 
$\exp\left(-S_{0}[\phi, {\wit \phi}]\right)$. 
Expanding the sine and using the Newton's binomial identity, we have
\BEA
\rho(t,s;\vec{r},\vec{r}') &=& \sum_{k=0}^{\infty}\sum_{p=0}^{2k+1} 
\rho_{k,p}(t,s;\vec{r},\vec{r}')
\nonumber \\
\rho_{k,p}(t,s;\vec{r},\vec{r}') &=& \frac{(-1)^{k+p+1}}{p!(2k+1-p)!} 
\left\langle {\wit \phi}(s,\vec{r}')\phi^{p}(t,\vec{r})
\phi^{2k+1-p}(s,\vec{r}')\exp -s[\phi, {\wit \phi}]\right\rangle_{0}  
\EEA
so that $\rho_{k,p}(t,s;\vec{r},\vec{r}')$ is given by 
a $2k+2$-point function, containing only one 
${\wit \phi}$ contribution and $2k+1$ fields $\phi$. The Bargman superselection
rule (\ref{A:B}) implies that the only 
contractions that will lead to non-vanishing averages come from the 
$k$-th order in the expansion of the exponential, so that we have
\BEQ
\rho_{k,p}(t,s;\vec{r},\vec{r}')=\frac{(-1)^{p+1}}{p!(2k+1-p)!k!}  
\int \prod_{j=1}^{k}\D j\D j'\: \left\langle{\wit \phi}(s,\vec{r}')\phi^{p}
(\vec{x},t)\phi^{2k+1-p}(s,\vec{r}'){\wit \phi(j)}\kappa(j,j'){\wit \phi(j')}
\right\rangle_{0} 
\EEQ   
where for clarity the notation $j$, (respectively $j'$) stands for 
$(u_{j},\vec{r}_{j})$, (respectively $(u_{j}^{'},\vec{r}_{j}^{'})$) 
and where integrals run over this set of $2k$ variables.  

Wick's theorem states that the integrand decomposes into sums of products 
of two-point functions. In order for a contraction not to vanish, the field 
${\wit \phi}(s,\vec{r}')$ must contract with one of the 
$p$ fields $\phi(t,\vec{r})$, which leads to 
\BEA
\rho_{k,0}(t,s;\vec{r},\vec{r}') &=& 0
\nonumber \\
\rho_{k,p}(t,s;\vec{r},\vec{r}' ) &=& \frac{(-1)^{p+1}}{(p-1)!(2k+1-p)!k!}
\int \prod_{j=1}^{k}\D j\D j'\: R_{0}(t,s;\vec{r},\vec{r}')
\nonumber \\
& & \times 
\left\langle
\phi^{p-1}(t,\vec{r})\phi^{2k+1-p}(s,\vec{r}')
{\wit \phi(j)}\kappa(j,j'){\wit \phi(j')}\right\rangle_{0} 
\EEA  
Summing over $p$, the response function $\rho(t,s;\vec{r},\vec{r}')$ reads
\BEQ
\rho(t,s;\vec{r},\vec{r}')=R_{0}(t,s;\vec{r},\vec{r}')\sum_{k=0}^{\infty} 
\frac{1}{(2k)!k!} \int \prod_{j=1}^{k}\D j\D j'\: 
\left\langle(\phi(t,\vec{r})-\phi(s,\vec{r}'))^{2k}
{\wit \phi(j)}\kappa(j,j'){\wit \phi(j')}\right\rangle_{0} 
\label{eqappA7}
\EEQ

At this stage, it will be interesting to give another equivalent expression 
of the magnetic correlation function $\Gamma(t,s;\vec{r},\vec{r}')$. 
Using the same strategy as before, but now expanding the cosine we find
\BEA
\Gamma(t,s;\vec{r},\vec{r}') &=& \sum_{k=0}^{\infty} 
\gamma_{k}(t,s;\vec{r},\vec{r}')
\nonumber \\
\gamma_{k}(t,s;\vec{r},\vec{r}') &=& \frac{(-1)^{k}}{(2k)!} 
\left\langle  (\phi(t,\vec{r})-\phi(s,\vec{r}'))^{2k}
\exp -s[\phi, {\wit \phi}]\right\rangle_{0} 
\EEA
For the same reason as before, because of mass conservation, only the 
expansion of order $k$ of the exponential term will contribute and lead 
to non-vanishing contractions, such that 
\BEQ
\gamma_{k}(t,s;\vec{r},\vec{r}')=\frac{1}{(2k)!k!} 
\int \prod_{j=1}^{k}\D j\D j'\: \left\langle(\phi(t,\vec{r})
-\phi(s,\vec{r}'))^{2k}{\wit \phi(j)}\kappa(j,j')
{\wit \phi(j')}\right\rangle_{0} 
\label{eqappA8}
\EEQ
Comparing (\ref{eqappA7}) with (\ref{eqappA8}), we thus have 
\BEQ
\rho(t,s;\vec{r},\vec{r}')=R_{0}(t,s;\vec{r},\vec{r}')
\Gamma(t,s;\vec{r},\vec{r}')
\EEQ
This is the main result of this appendix and gives 
eq.~(\ref{eqXYres2}) in the text. 

%%%%%%%%%%%%%%%%%%%%%%%%%%%%%%%%%%%%%%%%%%%%%%%%%%%%%%%%%%%%%%%%%%%%%%%%%%%%%%%%
\appsection{B}{Scaling form of a special four-point function}
%%%%%%%%%%%%%%%%%%%%%%%%%%%%%%%%%%%%%%%%%%%%%%%%%%%%%%%%%%%%%%%%%%%%%%%%%%%%%%%%

We study the scaling of the four-point function
\BEQ
{\cal F} := R_0^{(4)}(t,s,0,0;\vec{0},\vec{0},\vec{R},\vec{R}') = 
\left\langle \phi(t,\vec{0})\phi(s,\vec{0}) 
\wit{\phi}(0,\vec{R})\wit{\phi}(0,\vec{R}')\right\rangle_0
\EEQ
of a theory in MSR formulation of which the noiseless part is 
Schr\"odinger-invariant. The field $\phi$ is assumed to be quasiprimary
with scaling dimension $x$ and mass $\cal M$ and the response field 
$\wit{\phi}$ should also be quasiprimary with scaling dimension $\wit{x}$
and mass $\wit{\cal M}=-{\cal M}$. 
The covariance conditions on $\cal R$ are the following (we use $v(t)=0$)
\BEA
\left( t\partial_t + s\partial_s +\frac{1}{2}\vec{R}\partial_{\vec{R}}
+\frac{1}{2}\vec{R}'\partial_{\vec{R}'} + (x+\wit{x}) \right) {\cal F} &=& 0
\nonumber \\
\left( t^2\partial_t + s^2\partial_s +(t+s)x 
-\frac{\cal M}{2}\left(\vec{R}^2+{\vec{R}'}^2\right) \right) {\cal F} &=& 0
\EEA
We shall not discuss here the general solution of these equations. For our
purposes it is enough to observe that if we decompose $\cal F$ in the 
following symmetrized way
\BEQ \label{C:Fgg}
{\cal F} = {\cal G}(t,\vec{R}) {\cal G}(s,\vec{R}') 
+ {\cal G}(t,\vec{R}') {\cal G}(s,\vec{R})
\EEQ
which for a free-field theory would follow from Wick's theorem, then
\BEQ \label{C:G}
{\cal G}(t,\vec{R}) = {\cal G}_0 t^{-x-\wit{x}} \exp\left[ 
-\frac{\cal M}{2}\frac{\vec{R}^2}{t}\right]
\EEQ
produces a solution to both covariance conditions. 

We apply this result to the preparation part $C_{pr}$ of the autocorrelation
function, with the intention to use the result in the spin-wave 
approximation of the XY model. From eq.~(\ref{eqC4}) we have 
\BEA
C_{pr} &=& \frac{1}{2} \int_{\mathbb{R}^{2d}}\!\D\vec{R}\D\vec{R}'\: 
a(\vec{R}-\vec{R}') R_0^{(4)}(t,s,0,0;\vec{0},\vec{0},\vec{R},\vec{R}')
\label{C:s1} \\
&=& \frac{T_i}{(2\pi)^d} \int_{\mathbb{R}^{3d}}\!\D\vec{R}\D\vec{R}'\D\vec{q}\:
\frac{(ts)^{-(x+\wit{x})/2}}{\vec{q}^2} 
\exp\left[ \II\vec{q}\cdot(\vec{R}-\vec{R}')-\frac{\cal M}{2}\left(
\frac{\vec{R}^2}{t}+\frac{\vec{{R}'}^2}{s}\right)\right] 
\label{C:s2} \\
&=& \int_{0}^{\infty} \!\D z \int_{\mathbb{R}^{d}}\frac{\D\vec{q}}{(2\pi)^d} 
T_i e^{-\vec{q}^2(t+s+z)}
\label{C:s3} \\
&=& \frac{(4\pi)^{-d/2}}{d/2-1} T_i (t+s)^{1-d/2}
\label{C:s4}
\EEA 
where in going to (\ref{C:s2}) we used the initial condition (\ref{eqiniXY}) 
and the result (\ref{C:Fgg},\ref{C:G}) from above, 
next in going to (\ref{C:s3}) we
specialized to a free Gaussian field (where $x=\wit{x}=d$) of mass
${\cal M}=1/2$ and the last step we also assumed $d>2$. 

Eq.~(\ref{C:s4}) provides the preparation term, for any initial temperature
$T_i$, as required for the analysis of the autocorrelation function in the
XY model in section~5.2

%%%%%%%%%%%%%%%%%%%%%%%%%%%%%%%%%%%%%%%%%%%%%%%%%%%%%%%%%%%%%%%%%%%%%%%%%%%%%%%%
\appsection{C}{On local scale-invariance in the $1D$ Glauber-Ising model}
%%%%%%%%%%%%%%%%%%%%%%%%%%%%%%%%%%%%%%%%%%%%%%%%%%%%%%%%%%%%%%%%%%%%%%%%%%%%%%%%
\subsection{Two-point functions in the $1D$ Glauber-Ising model}

In the text we have seen that local scale-invariance implies the following
form of the two-time autoresponse function
\BEQ
R(t,s)=r_{0}\left(t-s\right)^{-1-a}\left(\frac{t}{s}\right)^{1+a-\lambda_{R}/z}
\label{eqappB1}
\EEQ    
In spite of a nice agreement with a large variety of models, this expression is 
not verified for the $1D$ Ising model with Glauber dynamics at $T=0$. 

The $1D$ Ising model is described by spins $\sigma_i=\pm 1$ and the 
Hamiltonian 
\BEQ
{\cal H} = - \sum_{i=1} \sigma_i \sigma_{i+1}
\EEQ
The exactly solvable Glauber dynamics \cite{Glau63} may be given through
the heat-bath rule, which gives the probability of finding the spin variables
$\sigma_{i}(t+1)$ in terms of those at time $t$
\BEQ
P\left(\sigma_i(t+1) = \pm 1\right) = \frac{1}{2} \left[
1\pm \tanh\left( \beta \left(\sigma_{i-1}(t)+
\sigma_{i+1}(t)+h_{i}(t)\right) \right) \right] 
\EEQ
where $\beta=1/T$ is the inverse temperature and $h_{i}(t)$ the external
magnetic field. Ageing occurs in this model at $T=0$. In the long-time
scaling limit, the two-time autocorrelation and autoresponse functions are
\cite{Godr00a,Lipp00,Maye03} 
\BEA
R(t,s) &=& \left.\frac{\delta \langle\sigma_i(t)\rangle}{\delta
h_i(s)}\right|_{h=0} \:=\: \frac{1}{\pi \sqrt{2s(t-s)\:}} 
\label{eqIsingR} \\
C(t,s) &=& \langle \sigma_i(t)\sigma_i(s)\rangle \:=\: 
\frac{2}{\pi}\arctan \sqrt{\frac{2}{t/s-1}\:} 
\label{eqIsingC}
\EEA
While these results were obtained first for a fully disordered initial state,
they remain true for long-ranged initial conditions
$\langle\sigma_r(0)\sigma_0(0)\rangle\sim r^{-\nu}$ with $\nu>0$ \cite{Henk03d}.
The case $\nu=0$ corresponds to the case of an initial magnetization $m_0$. Then
the connected part of $C(t,s)$ as well as $R(t,s)$ are multiplied by $1-m_0^2$
\cite{Pico02,Henk03d}. In any case, the forms of the scaling functions 
$f_{C,R}(y)$ are unchanged. 

Although these two-point functions clearly display dynamical scaling, it is
evident that the scaling form of $R(t,s)$ from (\ref{eqIsingR}) is incompatible
with the form suggested in (\ref{eqappB1}). Local scale-invariance as
developed in the text does not hold in the $1D$ Glauber-Ising model.

\subsection{Generalized realization of the ageing algebra}

We now show how Schr\"odinger invariance can be generalized such that the
exact response function (\ref{eqIsingR}) can be reproduced. Obviously,
time-translation invariance is broken in ageing systems. Therefore, as
already pointed out in \cite{Henk01,Henk02}, the dynamical symmetry 
cannot be the Schr\"odinger Lie algebra $\mathfrak{sch}_1$ which contains the 
time-translation generator $X_{-1}=-\partial_t$, but a subalgebra without
this generator might be acceptable. We consider the algebra \cite{Henk03}
\BEQ
\mathfrak{age}_1 := \{ X_0, X_1, Y_{-1/2}, Y_{1/2}, M_0 \}
\EEQ
and keeping the commutation relations (\ref{2:gl:SCHcomm}) we now look for
a more general realization of $\mathfrak{age}_1$. In this way, we write the
generators as $\{\Xi_{0,1},\Upsilon_{\pm 1/2},M_0\}$. These must be of the form
\BEA 
\Xi_{0} &=& -t\partial_t -\frac{1}{2}r\partial_r -\frac{x}{2} 
\nonumber \\
\Xi_{1} &=& -t^2\partial_t -tr\partial_r -xt -g(t) - \frac{\cal M}{2}r^2 
\nonumber \\
\Upsilon_{-1/2} &=& -\partial_r 
\nonumber \\
\Upsilon_{1/2} &=& - t\partial_r - {\cal M} r 
\nonumber \\
M_0 &=& - {\cal M}
\label{eqappB3}
\EEA
where $g=g(t)$ is to be found. 
The only commutator of $\mathfrak{age}_1$ constraining $g$ is 
\BEQ
\left[ \Xi_{1}, \Xi_{0} \right]=\Xi_{1}
\label{eqappB4}
\EEQ
which leads to 
\BEQ
t\partial_t g-g=0
\label{eqappB5}
\EEQ 
with the solution $g(t)=Kt$, with $K$ some constant. From (\ref{eqappB3}), the 
dynamical exponent $z=2$. If we were to require
in addition $\left[\Xi_{1},\Xi_{-1}\right]=2\Xi_{0}$ (and thereby go from
$\mathfrak{age}_1$ back to $\mathfrak{sch}_1$) we would recover $K=0$. 
Now, a quasiprimary field $\phi$ of $\mathfrak{age}_1$ will be 
characterized by a triplett $(x,K,{\cal M})$. 

We can now generalize local scale-invariance by requiring that the autoresponse 
function $R(t,s)$ formed from a quasiprimary field $\phi$ and its associated 
quasiprimary response field $\wit{\phi}$ to transform covariantly under the 
generators $\Xi_{0}$ and $\Xi_{1}$. It is a 
solution of the system of linear partial differential equations
\BEA
\left[ t\partial_t+\frac{x}{2}+s\partial_s+\frac{{\wit x}}{2} \right] R(t,s) 
&=& 0 
\nonumber \\
\left[ t^2\partial_t+(K+x)t+s^2\partial_s+({\wit K}+{\wit x})s\right] R(t,s) 
&=& 0
\label{eqappB6}
\EEA    
where ${\wit x}$ and ${\wit K}$ refer to the response field $\wit{\phi}$.  
Solving the system (\ref{eqappB6}) gives as final result
\BEA
R(t,s) &=& s^{-1-a}f_{R}(t/s)
\nonumber \\
f_{R}(y) &=& r_{0}y^{1+A-\lambda_{R}/2}\left(y-1\right)^{-1-A}
\label{eqappB7}
\EEA   
where the three independent non-equilibrium exponents $a,A,\lambda_R$ are 
\BEA
a &=& \frac{x+{\wit x}}{2}-1
\nonumber \\
A &=& a+K+{\wit K}
\nonumber \\
\lambda_{R} &=& 2x+2K
\label{eqappB8}
\EEA
In contrast with the previous realization of $\mathfrak{age}_1$, 
$K,\wit{K}\ne 0$ is possible and then $a$ and $A$ differ from each other. 

Comparison with the exact result (\ref{eqIsingR}) of the $1D$ Glauber-Ising 
model now gives complete agreement and we identify the exponents
\BEQ
a = 0 \;\; , \;\; A = -\frac{1}{2} \;\; , \;\; \lambda_R = 1
\EEQ
Of course, the values of $a$ and $\lambda_R$ have been obtained before
\cite{Godr00a,Lipp00} but $A$ seems to be a new exponent. 

At present, it must remain open whether the unusual properties of the $1D$
Glauber-Ising model are related to the fact that $T_c=0$ and therefore the
critical and low-temperature properties might have become mixed. 

Also, it remains to be seen whether the form of the autocorrelation function
can be understood from the generalized realization of the ageing algebra
$\mathfrak{age}_1$. We hope to come back to this elsewhere.

\newpage

%%++++++++++++++++++++++++++++++++++++++++++++++++++++++++++++++++++++++++++++++

%%%%%%%%%%%%%%%%%%%%%%%%%%%%%%%%%%%%%%%%%%%%%%%%%%%%%%%%%%%%%%%%%%%%%%%%%%%%%%%%

\end{document}